\documentclass[journal]{IEEEtran}
\usepackage{etex}
\usepackage{graphicx}
\usepackage{epstopdf}
\usepackage[table]{xcolor}

\usepackage{amssymb}
\usepackage{pbox}
\usepackage{listings}
\usepackage{dsfont}
\usepackage{pmboxdraw}
\usepackage{multirow}
\usepackage{multicol}
\usepackage{booktabs}
\usepackage{MnSymbol}
\usepackage{wasysym}

\usepackage{times}
\usepackage{textcomp}

\usepackage{tikz}
\usepackage{pgfplots}
 \usetikzlibrary{plotmarks}

\def\ie{i.\,e.,~}
\def\eg{e.\,g.,~}

\title{ Proficiency of Power Values for Load Disaggregation} 
\author{

	\IEEEauthorblockN{Manfred P\"ochacker, Dominik Egarter, Wilfried Elmenreich \\}
	\IEEEauthorblockA{
	 Institute of Networked and Embedded Systems / Lakeside Labs \\
	Alpen-Adria Universit\"at Klagenfurt, Austria \\
	   manfred.poechacker@aau.at, dominik.egarter@aau.at, wilfried.elmenreich@aau.at
	 }
	}

\makeatletter
\def\blfootnote{
	\xdef\@thefnmark{}\@footnotetext
	}
\makeatother

\begin{document}

\definecolor{mygray}{rgb}{0.5,0.5,0.5}
\definecolor{rot}{rgb}{0.9,0.3,0.1}
\definecolor{grun}{rgb}{0.1,0.9,0.3}
\definecolor{blau}{rgb}{0.3,0.1,0.9}

\maketitle

\blfootnote{ 
This work was performed in the research cluster Lakeside Labs funded by the European Regional Development Fund, the Carinthian Economic Promotion Fund (KWF), and the state of Austria under grants 20214/22935/34445 (Smart Microgrid Lab) and 20214/23743/35470 (Project MONERGY).
}

\begin{abstract}
Load disaggregation techniques infer the operation of different power consuming devices from a single measurement point that records the total power draw over time. Thus, a device consuming power at the moment can be understood as information encoded in the power draw. However, similar power draws or similar combinations of power draws limit the ability to detect the currently active device set. We present an information coding perspective of load disaggregation to enable a better understanding of this process and to support its future improvement. 
In typical cases of quantity and type of devices and their respective power consumption, not all possible device configurations can be mapped to distinguishable power values. We introduce the term of proficiency to describe the suitability of a device set for load disaggregation. We provide the notion and calculation of entropy of initial device states, mutual information of power values and the resulting uncertainty coefficient or proficiency. We show that the proficiency is highly dependent from the device running probability especially for devices with multiple states of power consumption. The application of the concept is demonstrated by exemplary artificial data as well as with actual power consumption data from real-world power draw datasets.
\end{abstract}



\section{Introduction}
\label{sec:Introduction}

There are several reasons why it is beneficial for a power grid to get as much information as possible in order to accomplish monitoring and controlling purposes, like giving consumption feedback, or detecting devices with high energy consumption. To avoid additional costs on hardware, installation and operation, it is highly valuable to derive this information from few, if not a single, measurement point(s).
Load disaggregation or Non Intrusive Load Monitoring (NILM) is a technique used for reasoning about the operation of power consuming devices from a single measurement point recording the total power draw. One of its promising applications is the field of metering within smart homes \cite{Peretto2010}, where information about single appliance usage is of high interest but monitoring with many sensors is not an option. Low cost power monitoring on device level is one step in integration of residential buildings into the future smart grid, which is considered to be a key-technology for carbon dioxide reduction.
NILM works based on information about the involved devices and permissible assumptions on usage scenarios. Replicable, as power consuming devices unintentionally encode information into the total power draw.
Load disaggregation algorithms identify attributes within the measured data and draw meaningful conclusions about the overall consumption scenario.

Load disaggregation that works exclusively based on (active) power values is of high interest because active power is simple to measure and existing metering infrastructure usually provide the necessary values. With the upcoming smart meters accessing  the data gets even easier. 	 
However, a main drawback is that devices with similar consumption characteristics are hard to distinguish and simultaneously running devices add up in power values. As a consequence the search space of possible power values at least doubles in size with each additional device. Devices with multiple values of power consumption additionally complicate the task.
A single power value can be either caused by different devices with similar characteristics or by aggregation of multiple less consuming devices. Which is why the distinction of all possible scenarios by  using exclusively power values is difficult.

Within this paper we discuss the problem of indistinguishable power values caused by different device configurations.
We use concepts of information theory to quantize the problem for a given device set by introducing the concept of \emph{proficiency} for load disaggregation.
It allows to compare the extent of the problem for different device sets more objectively. We do so by using real data of different measurement campaigns and houses with multi-state devices.  
We further investigate how proficiency is influenced by statistical operation probabilities of single devices and outline how the insights are useful for improvements of future NILM algorithms.

The goal of this work is to better understand the mapping of device configuration scenarios to power values. We identify this as a coding procedure for information communication. This knowledge is helpful for further improvement of load disaggregation, which is decoding in that context. 
The basic problem is related to measurement accuracy but has a different root. The two problems can be clearly separated that is why we use exclusively information theory for discrete sources and combinatorics. In that sense we complement other work on limits of NILM by measurement accuracy as well as on quantification of disaggregation complexity for appliance sets.

Section \ref{sec:NILMasCommunicationProblem} is dedicated to explain NILM as an information communication problem.
Within section \ref{sec:stateSpace} the concepts of information theory are applied to the case of aggregated power values. Two exemplary device sets are introduced which contain solely on-off devices. By section \ref{sec:multiStates} these concepts are extended to the more general case of devices with multiply power consumption values.
In section \ref{sec:deviceSets} we apply the concepts to nine different appliance sets and compare them among themselves before we discuss the results and sketch possible follow up work in section \ref{sec:Discussion}. 
In the last section we summarize the results.


\section{Load Disaggregation within Information Communication}
\label{sec:NILMasCommunicationProblem}

\begin{figure}[htbp]
	\centering
		\includegraphics[width=.48\textwidth]{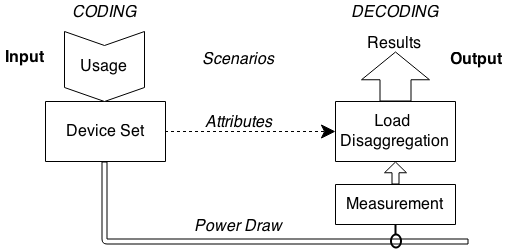}
	\caption{Load disaggregation is the decoding procedure in an information communication process. 
	}
	\label{fig:LoadDisaggregation}
\end{figure}

Figure \ref{fig:LoadDisaggregation} shows a scheme of information communication applied to load disaggregation according to how it was introduced in \cite{Hart1992} by Hart. It identifies Load Disaggregation as a decoding problem in the context of information communication theory. 
Load monitoring benefits from being non intrusive which means that any installation or device marking system is avoided.   
The primary source of information is the appliance usage which causes power consumption.
As the main purpose of a power cable is power supply, the utilized information content is produced unintentionally.
The meaningful decoding of a signal stream on the power cable is the challenge of load disaggregation. 
The code, which is the mapping of the usage scenarios to the power line signals, is exclusively defined by the devices and their attributes. 
There are various attributes that enable identification of a specific device by its fingerprint on the power cable. Frequency and non-harmonic device feedback on the input current is rich on information but the required high resolution measurement is usually costly and transmission functions of the power line circuits and their influences are not known. 
To overcome this and for additional arguments provided by 
This is one reason why Hart \cite{Hart1992} recommends usage of so called steady state attributes like power values for device detection.   

Several research has been done on the process shown in figure \ref{fig:LoadDisaggregation}. 
Dong analyzes in \cite{Dong2013a} limits for scenario detection due to measurement accuracy. 
A successful and efficient detection of the desired scenario requests the different parts to be well coordinated. 
The applications differ significantly concerning the acceptable effort on accuracy, maintenance, installation, computing power, measurement and finally costs. 
There are examples that very high measurement rates enable distinguishing quite specific scenarios, \eg which channel is watched on TV. 
But for most applications a very high data volume causes more burden than benefit. 

Current approaches solving the load disaggregation problem can be divided into between supervised and unsupervised approaches. A good overview on supervised approaches is given in \cite{Zeifman2011} and \cite{Zoha2012}.
The supervised approach needs a labeled data set to train a classifier and can be divided into optimization and pattern recognition \cite{Shaw2008} based algorithms.
In the optimization based approaches, the problem of aggregated power profiles is modelled as an optimization problem.
The total power consumption and a database of known power profiles of appliances are given. With this knowledge, a random composition of database power profiles is selected to estimate the total power consumption with minimal error \cite{Liang2010,Baranski2004}.
In pattern recognition approaches, proposed methods can be divided into clustering approaches \cite{Hart1992}, neural networks algorithms \cite{Srinivasan2006,Bier2012,Xu2015} and support vector machines based algorithms \cite{Kramer2012,Altrabalsi2014}.
The disadvantage of the supervised classification approaches is the requirement of  \textit{a priori} information.

Accordingly, recent research is more concerned with unsupervised algorithms, which are using unlabeled data.
Unsupervised algorithms do not require any training data and therefore no {\em a priori} information of the system.
Current approaches are based on dynamic time warping \cite{Liao2014}, clustering with blind source separation \cite{gonccalves2011unsupervised},  Hidden Markov Models (HMM) \cite{Zia2011,pattem2012,Egarter2014}, Fractorial HMM \cite{Zoha2013} other variations of HMM \cite{Zico2012,Kim2011}, temporal motif mining \cite{Shao2013} and blind source separation \cite{goncalves_unsupervised_2011}.
For all of these approaches the distinction between appliances is unsupervised whereas the labeling of a model with the corresponding appliance is not done automatically.
Approaches performing automatic labeling are conducted based on Bayesian inference \cite{Johnson2013} and semi-supervised classification \cite{Parson2014}.

The device states (specifically their values for power consumption) define the set of all possible device configurations, i.e., the state space. 
The usage is unknown and generates the aggregated power draw. Usage is dependent from the device operators and the build-in programs which make devices change their power consumption. The usage maps the possible device states to power-values. 
Load disaggregation is reversing what usage does: The power profile constitutes input and the current device states can be derived. 
The mapping of device states to power values is a coding process. The code depends on the power values of the device states, exclusively. 
There is no guarantee that this code is uniquely decodable and possibilities to modify that code are limited.

Additional difficulties arise in the practice of load disaggregation, \eg measurement resolution or noise, are not considered within this work. 
The theoretical constraints demonstrated within this paper arise for an idealized case where integer power values characterize the device states. 
The explicit inclusion of  measurement accuracy is on the one hand not necessary to demonstrate what we aim for and on the other hand offers no solution of the problem. 
The basic concepts are elucidated with on-off devices are extended to multi-state devices, subsequently. 
Correlation between different states and time durations are not taken into account.


\section{The state space of an appliance set}
\label{sec:stateSpace}

Within this section only on-off devices with only one single value of power consumption $P^d$ are considered.
The set of devices and so the set of power values
 \[ \mathbf{P}^D:\{P_1, ... , P_N\} \] 
is known.  We define the order of the device set in a way that $P_d<P_{d+1}$.
Without any additional knowledge it is possible to calculate all the possible power values $P_k$ by aggregation.
The state number $k$ specifies the subset of devices which is turned on and the complementary subset which is turned off. 
The first state is defined as the power value $P_1=0$ and the last states power value is the sum of all single devices 
\begin{equation}
P_M=P_{total}=\sum_{d=1}^N{P_d}
\label{eq:Ptotal}
\end{equation} 
which can be used to characterize the device set. 
\begin{table}[h]%
\small
\begin{tabular}{c|ccccccc}
\toprule
$k$	&   1 & 2\dots N+1 & \dots  & \dots & \dots & M-N\dots M-1 & M\\
$z$ & 0 & 1 & 2 & \dots  & N-2 & N-1 & N  \\
$n_z$ & 1 & N & ${N\choose 2}$ & \dots & ${N\choose {N-2}}$ & N & 1   \\
\bottomrule
\end{tabular}
\caption{The table enumerates the $M$ power states,  the number of turned on appliances $z$ and $n_z$, the number of different states with the same $z$. }
\label{tab:state}
\end{table}
\normalsize
In between these particular cases are always ${N\choose z}$ cases where $z$ out of the $N$ devices are turned on. The total number of possible states results to 
	\begin{equation}
		M = \sum^N_{z=0}{N\choose z} = 2^N
	\end{equation}
which is equal to the possible states of a binary word of length $N$. 
In the context of load disaggregation some of those states are very unlikely, even practicable impossible, to occur. But {\em a priori}, without knowing anything about the source and the emitted load profile it is impossible to detect which ones are more likely to occur.

How these $M$ states map to power values depends on the properties of the device set, i.e., the single device power values. 
The power value $P_k$ of a specific state $k$ is calculated by
\begin{equation}
 P_k= \sum_{d=1}^N{ S_{kd} P_d  } 
\label{eq:Pk}
\end{equation}
where $S_{kd}$ is the state matrix that contains a vector for each state $k$ that holds a $1$ for turned-on and a $0$ for turned-off devices. 
Repetition for all the states leads to the set of possible aggregated power values. 

Further we refer to two exemplary device sets, each containing ten on-off devices. 
The device set A has a linear power spectrum, in the sense that 
\begin{equation}
	P_d=P_{d-1}+P_{\Delta}
\label{eq:linSet}
\end{equation}
where we use $P_{1}=P_{\Delta}=5$W. 
Device set B contains the power values $\mathbf{P}^D:\{ 1, 2, 3, 5, 8, 14, 24, 41, 69, 117\}$. 
That power spectrum can be approximated by 
\begin{equation}
	P_d \approx  \alpha P_{d-1} 
\label{eq:expSet}
\end{equation}
for $\alpha=1.69$ and $P_{1}=1$W and therefore is of power law type.
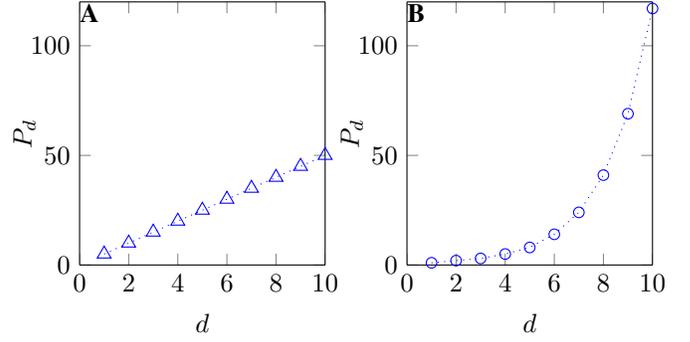
\begin{figure}[h]
	\centering
%
%
\begin{tikzpicture}

\begin{axis}[%
width=.18\textwidth,
height=3.5cm,
at={(.29\textwidth,0.8cm)},
scale only axis,
separate axis lines,
every outer x axis line/.append style={black},
every x tick label/.append style={font=\color{black}},
xmin=0,
xmax=10,
xlabel={$d$},
every outer y axis line/.append style={black},
every y tick label/.append style={font=\color{black}},
ymin=0,
ymax=120,
ylabel={${P}_{d}$},
ylabel style={yshift=-5mm} 
]
\addplot [color=blue,dotted,mark=o,mark options={solid},mark size=2.0pt,forget plot]
  table[row sep=crcr]{%
1	1\\
2	2\\
3	3\\
4	5\\
5	8\\
6	14\\
7	24\\
8	41\\
9	69\\
10	117\\
};
\node[below, right, align=left, inner sep=0mm, font=\bfseries, text=black] 
at (axis cs:0,115,0) { B};
\end{axis}

\begin{axis}[%
width=.18\textwidth,
height=3.5cm,
at={(.05\textwidth,0.8cm)},
scale only axis,
separate axis lines,
every outer x axis line/.append style={black},
every x tick label/.append style={font=\color{black}},
xmin=0,
xmax=10,
xlabel={$d$},
every outer y axis line/.append style={black},
every y tick label/.append style={font=\color{black}},
ymin=0,
ymax=120,
ylabel={${P}_{d}$},
ylabel style={yshift=-5mm} 
]
\addplot [color=blue,dotted,mark=triangle,mark options={solid},mark size=3.0pt,forget plot]
  table[row sep=crcr]{%
1	5\\
2	10\\
3	15\\
4	20\\
5	25\\
6	30\\
7	35\\
8	40\\
9	45\\
10	50\\
};
\node[below, right, align=left, inner sep=0mm, font=\bfseries, text=black]
at (axis cs:0,115,0) { A};
\end{axis}
\end{tikzpicture}%
	\caption{ The power values of device set A follows Equation \ref{eq:linSet} and has $P_{total}=275$. The set B has $P_{total}=284$ and single device values according to Equation \ref{eq:expSet}. }
	\label{fig:linDevSpec}
\end{figure}
Additionally these two sets have comparable total power of $275$ and $284$ Watt, which are the same magnitude.

A load profile is a stream of power values $P_i$ of length $n$.  
The total consumed energy is
	\begin{equation}
	E = \sum^n_{i=1}{P_i \Delta{t} }
	\end{equation}
where $\Delta t$ is the sampling time and the power values $P_i$ are averaged within a sampling duration. 
The average power of a load profile is
	\begin{equation}
	\hat{P}=\frac{E}{n \Delta{t}} \quad .
	\end{equation}
The power values $P_i$ result from the aggregation of power values of turned-on devices at time step $i$ so that 
\begin{equation}
P(i) = \sum_{s=1}^N{P_s S_s(i)} 
\label{eq:Pi}
\end{equation} 
where $N$ is the number of all devices and $S^{on}_s(i)$ is a boolean state function which is $1$ when the device $s$ is operated at the time $i$ and $0$ otherwise.

\subsection{Equal state probability}
\label{sec:EqualStateProbability}

When there is no knowledge of a source available it is common in information theory to assume the maximum entropy case, which means equal likelihood for all possible source symbols. 
 All of the state probabilities $p_k$ have the same value of $1/M$ and the entropy of the source, which is defined as
	\begin{equation}
		H=- \sum^M_{k=1} p_k ld(p_k)   \quad,
	\label{eq:entropy}
	\end{equation} 
has its highest possible value of $H_{max}=ld(M)$. The binary logarithm $\log_2$ is written as $ld$. 
$H_{max}$ is an upper bound for the entropy of a discrete memory-less, time-invariant source (DMS). 
The entropy $H_{max}$ of a load-source depends on the number of states or devices. 
As a first step it would allow to compare the difficulty of load disaggregation problems with different numbers of devices. Furthermore it is an upper bound for entropy of any load profile from this source. 
An equal distribution of power states results in equal average run-time for each single device. 
Table \ref{tab:state} shows that there are as many states with one device on as there are with one device off. This leads to the conclusion that if all the $M$ states are hypothetically visited one time each device is running exactly $M/2$ times, which is half of the total duration. 
Therefore we get the correlation 
\begin{equation}
	\frac{1}{M} \sum^M_{k=1}{P_k} = \frac{1}{2} P_{total}	
\label{eq:Pk=Ps}
\end{equation}
between the average state power and the aggregated power of the device set $P_{total}$. 

\begin{figure}[h]
	\centering
			\input{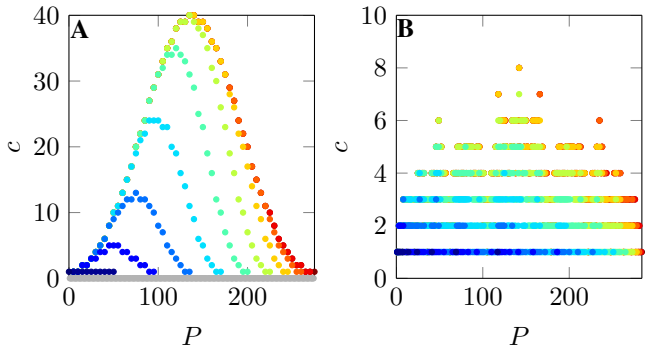}
	\caption{ Occupation numbers for set A and set B for all power values. The colors stand for the number of involved devices starting from $z=1$ in blue to $z=10$ in dark red.  }
	\label{fig:sfDevSpec}
\end{figure}

Counting the number of states with a specific power value $P_k$ gives a power state occupation number $c$.  
It can be written using the Dirac delta function as   
	\begin{equation}
	  c(P) =	\sum^M_{k=1} \int_0^\infty{  { \delta( P_k - P ) } dP}  \quad .
	\label{eq:defStatOcNum}
	\end{equation}
An occupation number above one reflects the challenge of distinguishing between different states consuming the same power. 
States with this power value are not uniquely distinguishable. 
Figure \ref{fig:sfDevSpec} shows the occupation numbers for the exemplary device sets which have the same total number of states.  
For set A there are up to forty states that map to the same power value while there are up to eight in set B. In set A a majority of power values (gray color) is not used at all.  
Load disaggregation is therefore expected to be more difficult for set A than for set B. 

The power values in figure \ref{fig:sfDevSpec} represent all states of a space between zero and $P_{total}$ which are available to encode the primary information. In that sense power values are the channel within a  theoretical communication setup where the device states are communicated. Otherwise as for the classical coding problem in communication theory the coding scheme is fixed and can not be designed according to the channel transmission function. 
The power values can be seen as the information source of the receiver side. 
In this context it´s entropy is the mutual information of the power value set $I^P$ and is calculated by
	\begin{equation}
		I^P = - \sum_{P_j=0}^{P_{total}}{p(P_j) ld(p(P_j))}   \quad .
	\label{eq:mutInf1}
	\end{equation}
We assume the power values $P_j$ to be a discrete set between $0$ and $P_{total}$ but the definition can be extended to continuous probability density functions. For equal state likely hood the power value probability is calculated by $c(P_j)$ so that we can define 
	\begin{equation}
	I^P_{max} = - \sum_{P_j=0}^{P_{total}}{ \frac{c(P_j)}{M} ld \bigg( \frac{c(P_j)}{M} \bigg) } 
	\label{eq:Ipmax}
	\end{equation}
which is the transported information by power values for the maximal entropy case. Note that it is not the theoretical maximum of transportable information by these power values. 
Therefore it needs the averaged power state occupation number  
\begin{equation}
	\hat{c}= \langle c(P_j)\rangle \quad .
\label{eq:c_hat}
\end{equation}
In average each of the $\frac{M}{\hat{c}}$ power states is occurring with probability of $\frac{\hat{c}}{M}$ which can be used further to approximate the mutual Information by 
\[ I^P_{max} \leq  H_{max} -ld (\hat{c}) \quad.
\]
When the mutual information is smaller than the entropy of the source it means that not all information can be transmitted and therefore the stream can not be decoded completely.
As a measure for that loss of information we suggest the uncertainty coefficient or proficiency which is defined in information theory \cite{Cover2005} as
\begin{equation}
	C = \frac{I^P}{H} 
\label{eq:proficiency}
\end{equation}
and is shown to be a meaningful performance matrix by \cite{White2008}. 
We name the proficiency for the maximal entropy case $C_{max}$  
\[
C_{max} = \frac{I^P_{max}}{H_{max}} \leq 1-\frac{ld (\hat{c})}{H_{max}} \quad .
\]
which is restricted by an upper bound using the average occupation number.  

Table \ref{tab:AB_IandC} shows the developed information measures for the exemplary device sets A and B. 
	\begin{table}[h]
	\small
	\centering
	\def\arraystretch{1.2}
		\begin{tabular}{r|ccc}
		\toprule
			& $I^P_{max}$	& $C_{max}$ & $\hat{c}$ \\
		\midrule
		Set A & 5.33 & 0.53 & 18.3 \\	
		Set B  & 8.04 & 0.80 & 3.6 \\ 
		\bottomrule
		\end{tabular}	
	\caption{The developed measures of average information for to the device sets A and B.} 
	\label{tab:AB_IandC}
	\end{table}

For another hypothetical device set B2, that is similar to B with $\alpha=2$ and $N=10$, the occupation number is $1$ for all the power values as shown in figure \ref{fig:msSets_ocNum} (just like the binary representation of natural numbers). 
An equal probability in state space maps to equal distribution of power values with $c=1$ which makes the mutual information reach the value of $H_{max}$.
It requires the power values space to be at least as big as the device state space to enable unique decoding. 
It means that only in the case $H_{max}=I^P$ full load disaggregation by exclusive use of power values is possible. 
The proficiency and the averaged occupation number are both one in this case.


\subsection{Equal device probability}
\label{sec:EqualDeviceProbability}

From the point of load disaggregation it is more suitable to deal with device probabilities than with probabilities for the combined power states. It is easier to relate devices to different user scenarios than to power states. 
The user-dependent devices follow behavioral patterns, \eg starting coffee machine after getting up. 
Automatic devices (like a fridge) are turned on regularly and therefore form a major part of the base load in a power draw.
For many types of devices characteristic operation probabilities can be estimated \cite{Kim2011}.
Even though their occurrence can vary, most of them are more likely to be switched off. 
For sizing of power lines (in a household) utilization factors are standard in engineering. The reasonable assumption is, that not all devices (or plugs) are used simultaneously which allows installation of power lines with smaller cross-section, which is more economic. Power factors are around $0.5$ for households, little higher for industry or commercial installations and they are expected to contain a safety buffer.

However, the state probabilities $p_k$ can be easily estimated in case the single device operation probabilities $p_d$ are known. 
From equation \ref{eq:Pk} for the state power the calculation of the state probability $p_k$ can be derived as 
\begin{equation}
		p_k= \prod^N_{d=1}{ \bigg(S_{kd} p_d  + (1-S_{kd}) (1-p_d)  \bigg) }
\label{eq:pk}
\end{equation}
assuming that the devices are statistically independent. 
Specific device probabilities do not fit the maximum entropy assumption. But as there is no {\em a prior} knowledge on $p_d$ we use the expectancy value 
\begin{equation}
\hat{p} =\langle p_d  \rangle 
\label{eq:pHat}
\end{equation}
for each device to demonstrate how the device sets entropy is influenced. 
{\em A  posterior} the average probability $\hat{p}$ for running any device can be calculated by
\begin{equation}
\hat{p} =\frac{E}{P_{total} n \Delta{t} }
\label{eq:pHat_E}
\end{equation}
using the energy $E$ of a load profile of length $n$. 
In case the single devices run-times $n_d$ are known even the 
 device operation probability can be estimated by 
\[p_d=\frac{n_d}{n} \quad.\]

\begin{figure}[htbp]
	\centering
%
%
\definecolor{mycolor1}{rgb}{0.00000,0.75000,0.75000}%
\definecolor{mycolor2}{rgb}{0.75000,0.00000,0.75000}%
\begin{tikzpicture}

\begin{axis}[%
width=.18\textwidth,
height=3.5cm,
at={(.29\textwidth,0.8cm)},
scale only axis,
separate axis lines,
every outer x axis line/.append style={black},
every x tick label/.append style={font=\color{black}},
xmin=0,
xmax=10,
xlabel={$z$},
every outer y axis line/.append style={black},
every y tick label/.append style={font=\color{black}},
ymin=0,
ymax=3,
ylabel={$h(z)$},
ylabel style={yshift=-5mm},
legend style={at={(-1.2,1.05)},anchor=south west,legend columns=5,legend cell align=left,align=left,draw=black}
]
\addplot [color=blue,dash pattern=on 1pt off 3pt on 3pt off 3pt,line width=1.0pt]
  table[row sep=crcr]{%
0	0.530002015127946\\
1	1.81698502207205\\
2	1.52253945811208\\
3	0.633062638574261\\
4	0.158472703422628\\
5	0.0258466525052199\\
6	0.00282996400192186\\
7	0.000207410758002895\\
8	9.79755257981299e-06\\
9	2.70444203526882e-07\\
10	3.32192809488736e-09\\
};
\addlegendentry{$\hat{p}=0.1$};

\addplot [color=black!50!green,dotted,line width=1.0pt]
  table[row sep=crcr]{%
0	0.145354185123806\\
1	0.770930337784622\\
2	1.77219160951331\\
3	2.35153028132146\\
4	2.00827404239935\\
5	1.1586344779582\\
6	0.458729394835568\\
7	0.123345492692797\\
8	0.0215918184813432\\
9	0.00222478611509398\\
10	0.00010256608136992\\
};
\addlegendentry{$0.3$};

\addplot [color=red,solid,line width=1.0pt]
  table[row sep=crcr]{%
0	0.009765625\\
1	0.09765625\\
2	0.439453125\\
3	1.171875\\
4	2.05078125\\
5	2.4609375\\
6	2.05078125\\
7	1.171875\\
8	0.439453125\\
9	0.09765625\\
10	0.009765625\\
};
\addlegendentry{$0.5$};

\addplot [color=mycolor1,dotted,line width=1.0pt]
  table[row sep=crcr]{%
0	0.00010256608136992\\
1	0.00222478611509399\\
2	0.0215918184813432\\
3	0.123345492692797\\
4	0.458729394835568\\
5	1.1586344779582\\
6	2.00827404239935\\
7	2.35153028132146\\
8	1.77219160951331\\
9	0.770930337784622\\
10	0.145354185123806\\
};
\addlegendentry{$0.7$};

\addplot [color=mycolor2,dash pattern=on 1pt off 3pt on 3pt off 3pt,line width=1.0pt]
  table[row sep=crcr]{%
0	3.32192809488736e-09\\
1	2.70444203526881e-07\\
2	9.79755257981297e-06\\
3	0.000207410758002894\\
4	0.00282996400192186\\
5	0.0258466525052199\\
6	0.158472703422628\\
7	0.63306263857426\\
8	1.52253945811208\\
9	1.81698502207205\\
10	0.530002015127946\\
};
\addlegendentry{$0.9$};

\end{axis}

\begin{axis}[
width=.18\textwidth,
height=3.5cm,
at={(.05\textwidth,0.8cm)},
scale only axis,
separate axis lines,
every outer x axis line/.append style={black},
every x tick label/.append style={font=\color{black}},
xmin=0,
xmax=10,
xlabel={$z$},
every outer y axis line/.append style={black},
every y tick label/.append style={font=\color{black}},
ymode=log,
ymin=1e-10,
ymax=1,
yminorticks=true,
ylabel={$p_k(z)$},
ylabel style={yshift=-2mm}
]
\addplot [color=blue,dash pattern=on 1pt off 3pt on 3pt off 3pt,line width=1.0pt,forget plot]
  table[row sep=crcr]{%
0	0.3486784401\\
1	0.0387420489\\
2	0.0043046721\\
3	0.0004782969\\
4	5.31441e-05\\
5	5.9049e-06\\
6	6.561e-07\\
7	7.29e-08\\
8	8.1e-09\\
9	9e-10\\
10	1e-10\\
};

\addplot [color=black!50!green,dotted,line width=1.0pt,forget plot]
  table[row sep=crcr]{%
0	0.0282475249\\
1	0.0121060821\\
2	0.0051883209\\
3	0.0022235661\\
4	0.0009529569\\
5	0.0004084101\\
6	0.0001750329\\
7	7.50141e-05\\
8	3.21489e-05\\
9	1.37781e-05\\
10	5.9049e-06\\
};

\addplot [color=red,solid,line width=1.0pt,forget plot]
  table[row sep=crcr]{%
0	0.0009765625\\
1	0.0009765625\\
2	0.0009765625\\
3	0.0009765625\\
4	0.0009765625\\
5	0.0009765625\\
6	0.0009765625\\
7	0.0009765625\\
8	0.0009765625\\
9	0.0009765625\\
10	0.0009765625\\
};

\addplot [color=mycolor1,dotted,line width=1.0pt,forget plot]
  table[row sep=crcr]{%
0	5.90490000000001e-06\\
1	1.37781e-05\\
2	3.21489e-05\\
3	7.50141000000001e-05\\
4	0.0001750329\\
5	0.0004084101\\
6	0.0009529569\\
7	0.0022235661\\
8	0.0051883209\\
9	0.0121060821\\
10	0.0282475249\\
};

\addplot [color=mycolor2,dash pattern=on 1pt off 3pt on 3pt off 3pt,line width=1.0pt,forget plot]
  table[row sep=crcr]{%
0	9.99999999999998e-11\\
1	8.99999999999998e-10\\
2	8.09999999999999e-09\\
3	7.28999999999999e-08\\
4	6.56099999999999e-07\\
5	5.90489999999999e-06\\
6	5.31441e-05\\
7	0.0004782969\\
8	0.0043046721\\
9	0.0387420489\\
10	0.3486784401\\
};

\end{axis}
\end{tikzpicture}%
	\caption{The single state probabilities $p_k(z)$ (for $z$ turned on devices) depend on the (averaged) device operation probabilities. The entropy $h(z)$ is additionally determined by the number of states. } 
	\label{fig:probability}
\end{figure}
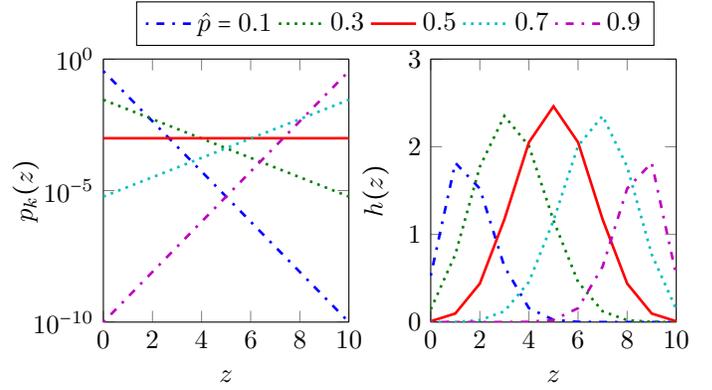

The average device probability is used to get 
\begin{equation}
p_k(z)= \hat{p}^{z(k)} (1-\hat{p})^{N-z(k)} 
\label{eq:pRho}
\end{equation} 
which is the state probability of a state with $z$ turned on devices. 
It is a logarithmic function as shown on the left hand side of figure \ref{fig:probability} for a set of ten devices. 
The state M, with all devices on, has the probability $\hat{p}^N $ and state $1$ has  $(1-\hat{p})^N $, respectively. 
The figure depicts  the entropy 
\[ h(z)=- {N\choose z} p_k(z) ln\big(p_k(z) \big) \] 
on the right hand side which is an intermediate result when calculating the total source entropy $	H = \sum_{z=1}^N{h(z)} $. 
In accordance with equation \ref{eq:Pk=Ps} the state probability is constant for the device probability of $\hat{p}=0.5$. 
The total source entropy, which is shown in Table \ref{tab:Hmax_p}, then reaches $H_{max}$.
	\begin{table}[h]
	\small
	\centering
	\def\arraystretch{1.2}
	\begin{tabular}{r|ccccc}
	\toprule
	$\hat{p}$ & 0.1	& 0.3 & 0.5 & 0.7 & 0.9 \\
	H & 4.69  &  8.81  & 10  &  8.81  &  4.69 \\
	\bottomrule
	\end{tabular}
	\caption{Total source entropy $H$ for different average device probabilities $\hat{p}$. } 
	\label{tab:Hmax_p}
	\end{table}
The entropy function $h(z)$ is symmetric with respect to $\hat{p}$ which means the total entropy for the operation probability of $0.1$  is the same as for the probability $0.1$ to be turned off.  

\begin{figure}[h]
	\centering
		\input{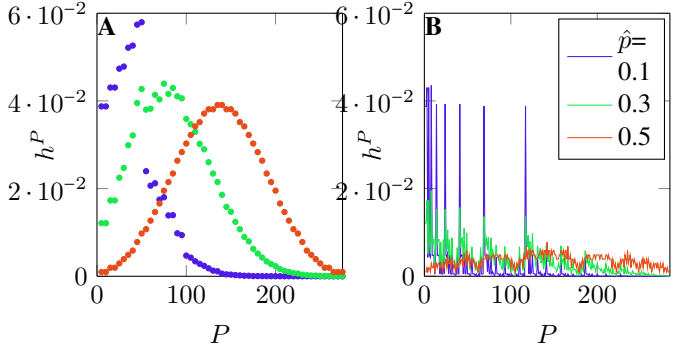}
	\caption{The power value probabilities determine the entropy function $h^P$ for the power states. We show it for set A and B for three different averaged device probabilities $\hat{p}$.   }
	\label{fig:pOcNo}
\end{figure}

The impact of device probabilities on the entropy propagates to power values, \ie mutual information and proficiency. 
The calculation of the power value probabilities  
	\begin{equation}
	  p(P) =	\sum^M_{k=1} p_k \int_0^\infty{  { \delta( P_k - P ) } dP}  \quad .
	\label{eq:defStatOcProb}
	\end{equation}
requires consideration of the state probability instead of merely the occupation number $c(P)$. 
This is used to calculate the mutual information
	\begin{equation}
		I^P = - \sum_{P_j=0}^{P_{total}} p(P_j) ld ( p(P_j) ) = \sum_{P_j=0}^{P_{total}} h^P(P_j)  
	\label{eq:mutInf}
	\end{equation}
of different single device probabilities.
The function $h^P(P_j)$ is shown in figure \ref{fig:pOcNo} for the exemplary device sets A and B. 
The three different values of $\hat{p}$ in figure \ref{fig:sfDevSpec} are used for calculating the mutual information $I^P$ and proficiency $C$ in table \ref{tab:AB_ICofp}.  
	\begin{table}[h]
	\small
	\centering
	\def\arraystretch{1.2}
	\begin{tabular}{c||c|ccc||c|ccc}
	\toprule
		 & $\hat{p}$&  0.1	& 0.3 & 0.5 & $\hat{p}$&  0.1	& 0.3 & 0.5\\
	\midrule
	  Set A & \multirow{2}{*}{$I^P$}& 3.70  &  5.14  &  5.33 & \multirow{2}{*}{$C$}& 0.79 & 0.58 & 0.53 \\
	  Set B &  &  4.50  &  7.51  &  8.04& & 0.96 & 0.85 & 0.80 \\
	\bottomrule
	\end{tabular}
	\caption{Mutual Information $I^P$ and Proficiency $C$ of the device sets A and B according to figure \ref{fig:pOcNo}. } 
	\label{tab:AB_ICofp}
	\end{table}
Even though the mutual information for $\hat{p}=0.5$ is higher, the proficiency and therefore the expected accuracy for disaggregation is lower than for $\hat{p}=0.1$. 
We conclude that equal probability for operation of all single devices does not lead to equal occurrence of the states or power values.


\section{Multi-State devices}
\label{sec:multiStates}

Multi-state appliances complicate the description of the power values for a set of devices. 
Power consumption for a device $d$ is specified by a vector with an entry for all its $\hat{s}_d$ power values. 
A device set with $N$ devices is, for instance, defined by
 \[ \mathbf{P}^D:\{ (P_1^1,P_1^2); (P_2^1,P_2^2,P_2^3);(P_3); \dots ; (P_N^1,\dots ,P_N^{\hat{s}_N})\} \quad.\] 
The second device has three power values $\hat{s}_2=3$, which means the device has four possible states. Device three is an on-off device with one power value. 
The total number of power values 
\begin{equation}
	S=\sum_{d=1}^N{ \hat{s}_d }
\label{eq:deviceStates}
\end{equation}
is a characteristic parameter for a device set (in case of exclusive on-off devices $S=N$). 
We assume that all power values are increasing in order to assure a unique description for a specific device set.  
The highest power value of a device  $P_d^{\hat{s}}$ defines the order within the device set so that  $P_d^{\hat{s}}<P_{d+1}^{\hat{s}}$. The power values of a single device are sorted that $P_d^s<P_d^{s+1}$.

Like in the case of simple devices the number of possible states is calculated by multiplication of the number of states for all the $N$ devices
\begin{equation}
	M=\prod_{d=1}^N{ ( \hat{s}_d+1 )}  \quad . 
\label{eq:Mms}
\end{equation}
The $M$ states map to the power values $P_k$  which can be calculated  by 
\begin{equation}
	P_k= \sum_{d=1}^N{ P_d^{S_{kd}}  } \quad .
\label{eq:Pk_multistates}
\end{equation}
using a different notation than in equation \ref{eq:Pk}.
The state matrix element $S_{kd}$ contains the power state of the device $d$ associated with state $k$, which is in accordance with its earlier usage.  
Now, $S_{kd}$ is used as an index not as an exponent, so 
 $P_d^{S_{kd}}$ is the power value of device $d$ associated with state $k$.
This notation requires the additional definition of $P^0$ for all devices in a way that 
\[  P_d^0=0 \quad \forall \quad d \in N \quad. \]
The mapping of the state number $k$ to the device power state is more difficult than for exclusive on-off devices but follows a straightforward principle. For the above example, device $2$ is off for the first $\hat{s}_1+1=3$ states, \ie $S_{1\dots3,2}=0$ or more generally $S_{1\dots3,d\geq2}=0$. For the states $k=4\dots6$ device $2$ runs with its first power value, \ie $S_{4\dots6,2}=1$ and so forth.
The power value of the last state is
\[ P_M= P_{total} =\sum_{d=1}^N{ P_d^{\hat{s}_d}}  \]
which is the highest possible one.
The occupation number $c$ is estimated from the set of power values according to (\ref{eq:defStatOcNum}).

The multiple possible device states require a modification of the state probabilities $p_k$. 
The device probability is written in the same way as the power values so that $p_d^s$ is the probability that device $d$ is running on power value $s$. The state probability is than gained  by
\begin{equation}
			p_k= \prod^N_{d=1}{ p_d^{S_{kd}}  }  
\label{eq:pk_ms}
\end{equation}
with usage of the device state matrix. 
The off-state probability $p_d^0$  needs to be calculated by 
\[ p_d^0=1-\sum_{s=1}^{\hat{s}_d}{p_d^s}  \]
as it is required within this notation.
The notation introduced for multi-state devices is more general and includes the two state devices from section \ref{sec:stateSpace}. 
The average device probability $\hat{p}$ is not an equal likelihood assumption. 
The assumption is on the likelihood of the off-states, the other device states are equally likely.  
The state probability is further used to estimate entropy (\ref{eq:entropy}), power values probability (\ref{eq:defStatOcProb}) and mutual information (\ref{eq:mutInf}) just as in the case of on-off devices.

\begin{figure}[h]
	\centering
		\resizebox {\columnwidth} {!} {
		\input{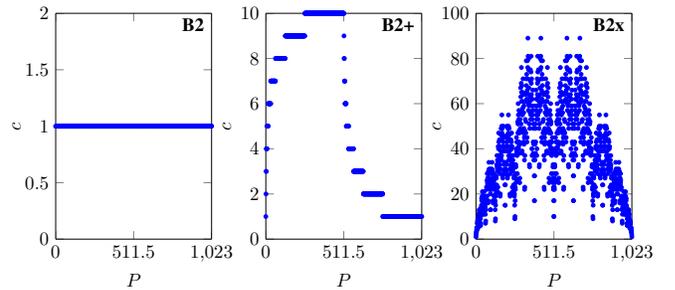}
		}
	\caption{Occupation numbers for power values of the three artificial device sets. }
	\label{fig:msSets_ocNum}
\end{figure}
To demonstrate the influence of multi-state devices we compare three artificial device sets. The multi-state device sets are based on the device set B2 which is constructed as set B but with the parameter $\alpha=2$, \ie all states have the occupation number $1$ what makes it trivial as visible in figure \ref{fig:msSets_ocNum}. 
Both derivative sets have $9$ additional states. For set B2+ device 10 has 9 additional states with the power values of devices 1 to 9. For set B2x the last nine devices have a second state with the power value of the previous device. 
\begin{table}[h]
	\small
	\center
	\def\arraystretch{1.2}

\begin{tabular}{c|cccccc}
		\toprule
	Device Set &	 $S$	&$M$ & $H_{max}$ & $I^P_{max}$ &$C_{max}$	& $\hat{c}$ \\ 
		\midrule
	\textbf{B}   & 10 & 1024 &10 &8.04&  0.80 & 3.6 \\
	\textbf{B2}  & 10 & 1024 &10 &10 & 1 &1 \\ 
	\textbf{B2+}  &  19 & 5632 &12.46 & 9.6 &0.77 &5.5 \\ 
	\textbf{B2x}  &  19 & 39366 &15.26&   9.8 & 0.64 & 38.5\\
	\bottomrule
		\end{tabular}	
	\caption{Several parameters for the artificial sets of ten devices. } 
	\label{tab:Bsets}
	\end{table}
The derivative sets have the same power values but they are differently distributed among the devices. 
The reference values for several artificial device sets are listed in table \ref{tab:Bsets}. Figure  \ref{fig:msSets_ocNum} shows the occupation numbers for the power values. 

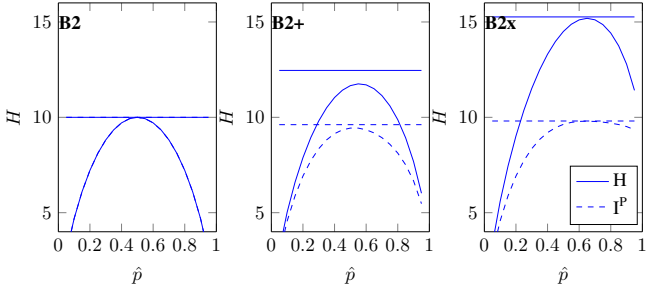
\begin{figure}[h]
	\centering
		\resizebox {\columnwidth} {!} {
%
%
\begin{tikzpicture}

\begin{axis}[%
width=1.15in,
height=1.666667in,
at={(3.6in,0.333333in)},
scale only axis,
separate axis lines,
every outer x axis line/.append style={black},
every x tick label/.append style={font=\color{black}},
xmin=0,
xmax=1,
xlabel={$\hat{p}$},
every outer y axis line/.append style={black},
every y tick label/.append style={font=\color{black}},
ymin=4,
ymax=16,
ylabel={$H$},
ylabel style={yshift=-4mm}, 
legend style={at={(0.97,0.03)},anchor=south east,legend cell align=left,align=left,draw=black}
]
\addplot [color=blue,solid]
  table[row sep=crcr]{%
0.05	3.31396957116135\\
0.1	5.58995593589429\\
0.15	7.44840304716267\\
0.2	9.01928094887567\\
0.25	10.3627812445946\\
0.3	11.5129089923084\\
0.35	12.4906805537517\\
0.4	13.3095059445504\\
0.45	13.9777445398739\\
0.5	14.5\\
0.55	14.8777445398828\\
0.6	15.1095059445528\\
0.65	15.1906805537596\\
0.7	15.1129089923051\\
0.75	14.8627812445951\\
0.8	14.4192809488712\\
0.85	13.7484030471695\\
0.9	12.7899559358918\\
0.95	11.4139695711598\\
};
\addlegendentry{H};

\addplot [color=blue,dashed]
  table[row sep=crcr]{%
0.05	2.817827783383\\
0.1	4.56845304492145\\
0.15	5.88819522511061\\
0.2	6.91747351172264\\
0.25	7.72574392396273\\
0.3	8.35621349445574\\
0.35	8.83961046789006\\
0.4	9.19996512728408\\
0.45	9.45749372746193\\
0.5	9.63017354827795\\
0.55	9.73457271808364\\
0.6	9.7860942185405\\
0.65	9.79860021084797\\
0.7	9.78339881530619\\
0.75	9.74793771917304\\
0.8	9.69481709176648\\
0.85	9.62098391146435\\
0.9	9.51557614125716\\
0.95	9.35136717533948\\
};
\addlegendentry{$\text{I}^\text{P}$};

\addplot [color=blue,solid,forget plot]
  table[row sep=crcr]{%
0.05	15.2646625064904\\
0.1	15.2646625064904\\
0.15	15.2646625064904\\
0.2	15.2646625064904\\
0.25	15.2646625064904\\
0.3	15.2646625064904\\
0.35	15.2646625064904\\
0.4	15.2646625064904\\
0.45	15.2646625064904\\
0.5	15.2646625064904\\
0.55	15.2646625064904\\
0.6	15.2646625064904\\
0.65	15.2646625064904\\
0.7	15.2646625064904\\
0.75	15.2646625064904\\
0.8	15.2646625064904\\
0.85	15.2646625064904\\
0.9	15.2646625064904\\
0.95	15.2646625064904\\
};
\addplot [color=blue,dashed,forget plot]
  table[row sep=crcr]{%
0.05	9.80738709103356\\
0.1	9.80738709103356\\
0.15	9.80738709103356\\
0.2	9.80738709103356\\
0.25	9.80738709103356\\
0.3	9.80738709103356\\
0.35	9.80738709103356\\
0.4	9.80738709103356\\
0.45	9.80738709103356\\
0.5	9.80738709103356\\
0.55	9.80738709103356\\
0.6	9.80738709103356\\
0.65	9.80738709103356\\
0.7	9.80738709103356\\
0.75	9.80738709103356\\
0.8	9.80738709103356\\
0.85	9.80738709103356\\
0.9	9.80738709103356\\
0.95	9.80738709103356\\
};
\node[below, right, align=left, inner sep=0mm, font=\bfseries, text=black]
at (axis cs:0,15,0) {  B2x};
\end{axis}

\begin{axis}[%
width=1.15in,
height=1.666667in,
at={(2.05in,0.333333in)},
scale only axis,
separate axis lines,
every outer x axis line/.append style={black},
every x tick label/.append style={font=\color{black}},
xmin=0,
xmax=1,
xlabel={$\hat{p}$},
every outer y axis line/.append style={black},
every y tick label/.append style={font=\color{black}},
ymin=4,
ymax=16,
ylabel={$H$},
ylabel style={yshift=-4mm} 
]
\addplot [color=blue,solid,forget plot]
  table[row sep=crcr]{%
0.05	3.03006597590411\\
0.1	5.02214874538093\\
0.15	6.59669226139596\\
0.2	7.88366656785206\\
0.25	8.94326326831357\\
0.3	9.80948742077321\\
0.35	10.5033553869651\\
0.4	11.0382771825015\\
0.45	11.422612182577\\
0.5	11.6609640474417\\
0.55	11.7548049920658\\
0.6	11.7026628014792\\
0.65	11.4999338154315\\
0.7	11.1382586587282\\
0.75	10.6042273157569\\
0.8	9.8768234247838\\
0.85	8.92204192781784\\
0.9	7.67969122129148\\
0.95	6.0198012613027\\
};
\addplot [color=blue,dashed,forget plot]
  table[row sep=crcr]{%
0.05	2.80205440001893\\
0.1	4.55395177158674\\
0.15	5.88168148496041\\
0.2	6.91922048656389\\
0.25	7.73081644712704\\
0.3	8.35512858975671\\
0.35	8.81883803533221\\
0.4	9.1424213216671\\
0.45	9.34303897400279\\
0.5	9.43590285271272\\
0.55	9.43429605125989\\
0.6	9.34813283478301\\
0.65	9.1817022796937\\
0.7	8.93188595018989\\
0.75	8.58691933196057\\
0.8	8.12439902386856\\
0.85	7.5065262036008\\
0.9	6.66768439337517\\
0.95	5.47293135504381\\
};
\addplot [color=blue,solid,forget plot]
  table[row sep=crcr]{%
0.05	12.4594316186373\\
0.1	12.4594316186373\\
0.15	12.4594316186373\\
0.2	12.4594316186373\\
0.25	12.4594316186373\\
0.3	12.4594316186373\\
0.35	12.4594316186373\\
0.4	12.4594316186373\\
0.45	12.4594316186373\\
0.5	12.4594316186373\\
0.55	12.4594316186373\\
0.6	12.4594316186373\\
0.65	12.4594316186373\\
0.7	12.4594316186373\\
0.75	12.4594316186373\\
0.8	12.4594316186373\\
0.85	12.4594316186373\\
0.9	12.4594316186373\\
0.95	12.4594316186373\\
};
\addplot [color=blue,dashed,forget plot]
  table[row sep=crcr]{%
0.05	9.61501043299683\\
0.1	9.61501043299683\\
0.15	9.61501043299683\\
0.2	9.61501043299683\\
0.25	9.61501043299683\\
0.3	9.61501043299683\\
0.35	9.61501043299683\\
0.4	9.61501043299683\\
0.45	9.61501043299683\\
0.5	9.61501043299683\\
0.55	9.61501043299683\\
0.6	9.61501043299683\\
0.65	9.61501043299683\\
0.7	9.61501043299683\\
0.75	9.61501043299683\\
0.8	9.61501043299683\\
0.85	9.61501043299683\\
0.9	9.61501043299683\\
0.95	9.61501043299683\\
};

\node[ right, align=left, inner sep=0mm, font=\bfseries, text=black]
at (axis cs:0,15,0) {  B2+};
\end{axis}

\begin{axis}[%
width=1.15in,
height=1.666667in,
at={(0.5in,0.333333in)},
scale only axis,
separate axis lines,
every outer x axis line/.append style={black},
every x tick label/.append style={font=\color{black}},
xmin=0,
xmax=1,
xlabel={$\hat{p}$},
every outer y axis line/.append style={black},
every y tick label/.append style={font=\color{black}},
ymin=4,
ymax=16,
ylabel={$H$},
ylabel style={yshift=-4mm} 
]
\addplot [color=blue,solid,forget plot]
  table[row sep=crcr]{%
0.05	2.8639695711596\\
0.1	4.68995593589284\\
0.15	6.09840304716395\\
0.2	7.21928094887374\\
0.25	8.11278124459131\\
0.3	8.81290899230692\\
0.35	9.34068055375488\\
0.4	9.70950594454664\\
0.45	9.92774453987803\\
0.5	10\\
0.55	9.92774453987803\\
0.6	9.70950594454663\\
0.65	9.34068055375491\\
0.7	8.81290899230691\\
0.75	8.11278124459135\\
0.8	7.21928094887366\\
0.85	6.09840304716399\\
0.9	4.68995593589281\\
0.95	2.86396957115956\\
};
\addplot [color=blue,dashed,forget plot]
  table[row sep=crcr]{%
0.05	2.8639695711596\\
0.1	4.68995593589284\\
0.15	6.09840304716395\\
0.2	7.21928094887374\\
0.25	8.11278124459131\\
0.3	8.81290899230692\\
0.35	9.34068055375488\\
0.4	9.70950594454664\\
0.45	9.92774453987803\\
0.5	10\\
0.55	9.92774453987803\\
0.6	9.70950594454663\\
0.65	9.34068055375491\\
0.7	8.81290899230691\\
0.75	8.11278124459135\\
0.8	7.21928094887366\\
0.85	6.09840304716399\\
0.9	4.68995593589281\\
0.95	2.86396957115956\\
};
\addplot [color=blue,solid,forget plot]
  table[row sep=crcr]{%
0.05	10\\
0.1	10\\
0.15	10\\
0.2	10\\
0.25	10\\
0.3	10\\
0.35	10\\
0.4	10\\
0.45	10\\
0.5	10\\
0.55	10\\
0.6	10\\
0.65	10\\
0.7	10\\
0.75	10\\
0.8	10\\
0.85	10\\
0.9	10\\
0.95	10\\
};
\addplot [color=blue,dashed,forget plot]
  table[row sep=crcr]{%
0.05	10\\
0.1	10\\
0.15	10\\
0.2	10\\
0.25	10\\
0.3	10\\
0.35	10\\
0.4	10\\
0.45	10\\
0.5	10\\
0.55	10\\
0.6	10\\
0.65	10\\
0.7	10\\
0.75	10\\
0.8	10\\
0.85	10\\
0.9	10\\
0.95	10\\
};
\node[below, right, align=left, inner sep=0mm, font=\bfseries, text=black]
at (axis cs:0,15,0) {  B2};
\end{axis}
\end{tikzpicture}%
		}
	\caption{Entropy $H$ and mutual Information $I^P$ as a function of the device probability $\hat{p}$ for the three artificial device sets. The horizontal lines mark the values for the maximal entropy case.}
	\label{fig:Bsets_pd}
\end{figure}

Entropy and mutual information are shown for different device probabilities in figure \ref{fig:Bsets_pd} including the values for the maximal entropy case. The maximum of the entropy curve for set B2, which is reaching $H_{max}$, shifts due to the additional power values in the extended sets. In set B2x most devices (9 of 10) have two power values which is equal to three states. For nine devices the equal distribution of states is equivalent to the device probability of $\hat{p}={2}/{3}$ which is where the maximum occurs.
For set B2+ the entropy function does not reach $H_{max}$ (depicted as horizontal line).  This is due to differences in the number of power states between the devices. While device 10 reaches equal state distribution in $\hat{p}\approx0.9$ all other two-state devices reach it at $0.5$. In other words device 10 is involved in many of the possible states but is not operated more frequently to the same extent.
The additional states in the derivative sets significantly increase the entropy while the mutual information is actually decreasing. 
This is caused by the constant total power $P_{total}$ of the three device sets.


\section{ Case study on real device sets }
\label{sec:deviceSets}
We apply the measures developed within this paper to realistic device sets. 
We chose data sets frequently used for test cases within load disaggregation studies. 
Such as the GreenD \cite{greend}, the RedD \cite{kolter2011} and the Eco \cite{Beckel2014} dataset as used in \cite{Egarter2015,Figueiredo2013}.
\begin{table}[h]
			\begin{tabular}{c|p{6.7cm}}
		\toprule
Device Set   &	Power States \\ 
		\midrule
		
  \textbf{GreenD 1}&  [55 140 240], [1220], [60 148 470 570 1225 1265], [1790], [70 155 210 260 423 1898],  [40 1900]  \\
	\textbf{GreenD 2} &  [60], [80],  [850], [1580], [80 1725], [90 173 1910]\\
	\textbf{GreenD 3}&   [110 235 285 360], [120 1235], [55 125 540 882 1047 1220 1630], [70 2002],  [125 245 358 1998 2100], [70 160 2358 2550] \\
			\midrule
	\textbf{RedD 1} &   [200 420], [50 210 410 890 1115], [260 710 1440], [55 110 270 300 620 1405 1505], [1680 2478], [2705] \\
	\textbf{RedD 2} & [123], [410], [160 420], [130 210 770],  [1050],  [40 1718 1850] \\
	\textbf{RedD 3} & [100 400], [210 525 730], [40 365 900 1220 1520], [860 960 1285 1605], [120 540 1698], [2265]\\
			\midrule
	\textbf{Eco 1}&  [40], [72], [250 440 785], [50 1225], [1800], [90 180 250 365 2168]\\
	\textbf{Eco 2}&    [70], [55 175], [80 185], [50 310], [50 1840], [120 2132]\\
	\textbf{Eco 3}&   [100], [120], [130], [100 175 280],  [40 1365 1485], [67 190 280 445 650 785 1065 1545]\\
		\bottomrule
		\end{tabular}	
	\caption{ Power values of the device sets according to \cite{Egarter2015}. } 
	\label{tab:dataSetPvalues}
	\end{table}
To ensure comparability we use exactly the same six appliances for each house as quoted as submetered power values in \cite{Egarter2015}.
The power states of the appliance set were detected by an algorithm presented in there.
For further information, \eg how to extract appliance state information and the choice of appliances, we refer to \cite{Egarter2015}.

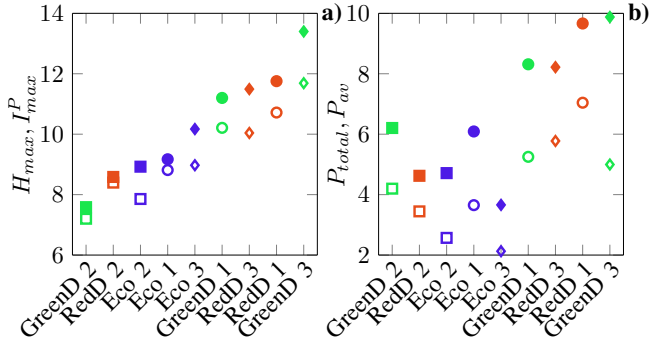
\begin{figure}[h]
	\centering
	\resizebox {\columnwidth}{!}{
%
%
\definecolor{mycolor1}{rgb}{0.10000,0.90000,0.30000}%
\definecolor{mycolor2}{rgb}{0.30000,0.10000,0.90000}%
\begin{tikzpicture}

\begin{axis}[%
width=.4\columnwidth,
height=3.5cm,
at={(.6\columnwidth,0.366667in)},
scale only axis,
clip=false,
separate axis lines,
every outer x axis line/.append style={black},
every x tick label/.append style={font=\color{black}},
xmin=0.5,
xmax=9.5,
xtick={1,2,3,4,5,6,7,8,9},
xticklabels={{GreenD 2},{RedD 2},{Eco 2},{Eco 1},{Eco 3},{GreenD 1},{RedD 3},{RedD 1},{GreenD 3}},
xticklabel style={
            inner sep=1pt,
            anchor=north east,
            rotate=45 },
every outer y axis line/.append style={black},
every y tick label/.append style={font=\color{black}},
ymin=2,
ymax=10,
ylabel={$P_{total},P_{av}$},
ylabel style={yshift=-6.5mm} 
]
\addplot [color=mycolor1,solid,line width=1.0pt,mark=square*,mark options={solid,fill=mycolor1},forget plot]
  table[row sep=crcr]{%
1	6.205\\
};
\addplot [color=mycolor1,solid,line width=1.0pt,mark=square,mark options={solid},forget plot]
  table[row sep=crcr]{%
1	4.1975\\
};
\addplot [color=orange!60!purple,solid,line width=1.0pt,mark=square*,mark options={solid,fill=orange!60!purple},forget plot]
  table[row sep=crcr]{%
2	4.623\\
};
\addplot [color=orange!60!purple,solid,line width=1.0pt,mark=square,mark options={solid},forget plot]
  table[row sep=crcr]{%
2	3.44566666666667\\
};
\addplot [color=mycolor2,solid,line width=1.0pt,mark=square*,mark options={solid,fill=mycolor2},forget plot]
  table[row sep=crcr]{%
3	4.712\\
};
\addplot [color=mycolor2,solid,line width=1.0pt,mark=square,mark options={solid},forget plot]
  table[row sep=crcr]{%
3	2.5685\\
};
\addplot [color=mycolor2,solid,line width=1.0pt,mark=*,mark options={solid,fill=mycolor2},forget plot]
  table[row sep=crcr]{%
4	6.09\\
};
\addplot [color=mycolor2,solid,line width=1.0pt,mark=o,mark options={solid},forget plot]
  table[row sep=crcr]{%
4	3.65176666666667\\
};
\addplot [color=mycolor2,solid,line width=1.0pt,mark=diamond*,mark options={solid,fill=mycolor2},forget plot]
  table[row sep=crcr]{%
5	3.66\\
};
\addplot [color=mycolor2,solid,line width=1.0pt,mark=diamond,mark options={solid},forget plot]
  table[row sep=crcr]{%
5	2.12670833333333\\
};
\addplot [color=mycolor1,solid,line width=1.0pt,mark=*,mark options={solid,fill=mycolor1},forget plot]
  table[row sep=crcr]{%
6	8.313\\
};
\addplot [color=mycolor1,solid,line width=1.0pt,mark=o,mark options={solid},forget plot]
  table[row sep=crcr]{%
6	5.25066666666667\\
};
\addplot [color=orange!60!purple,solid,line width=1.0pt,mark=diamond*,mark options={solid,fill=orange!60!purple},forget plot]
  table[row sep=crcr]{%
7	8.218\\
};
\addplot [color=orange!60!purple,solid,line width=1.0pt,mark=diamond,mark options={solid},forget plot]
  table[row sep=crcr]{%
7	5.77583333333333\\
};
\addplot [color=orange!60!purple,solid,line width=1.0pt,mark=*,mark options={solid,fill=orange!60!purple},forget plot]
  table[row sep=crcr]{%
8	9.663\\
};
\addplot [color=orange!60!purple,solid,line width=1.0pt,mark=o,mark options={solid},forget plot]
  table[row sep=crcr]{%
8	7.04161904761905\\
};
\addplot [color=mycolor1,solid,line width=1.0pt,mark=diamond*,mark options={solid,fill=mycolor1},forget plot]
  table[row sep=crcr]{%
9	9.877\\
};
\addplot [color=mycolor1,solid,line width=1.0pt,mark=diamond,mark options={solid},forget plot]
  table[row sep=crcr]{%
9	4.99627142857143\\
};
\node[right, align=left, inner sep=0mm, font=\bfseries, text=black]
at (axis cs:9.65,10,0) {b)};
\end{axis}

\begin{axis}[%
width=.4\columnwidth,
height=3.5cm,
at={(.1\columnwidth,0.366667in)},
scale only axis,
clip=false,
separate axis lines,
every outer x axis line/.append style={black},
every x tick label/.append style={font=\color{black}},
xmin=0.5,
xmax=9.5,
xtick={1,2,3,4,5,6,7,8,9},
xticklabels={{GreenD 2},{RedD 2},{Eco 2},{Eco 1},{Eco 3},{GreenD 1},{RedD 3},{RedD 1},{GreenD 3}},
xticklabel style={
            inner sep=1pt,
            anchor=north east,
            rotate=45 },
every outer y axis line/.append style={black},
every y tick label/.append style={font=\color{black}},
ymin=6,
ymax=14,
ylabel={$H_{max},I^{P}_{max}$},
ylabel style={yshift=-5mm} 
]
\addplot [color=mycolor1,solid,line width=1.0pt,mark=square*,mark options={solid,fill=mycolor1},forget plot]
  table[row sep=crcr]{%
1	7.58496250072116\\
};
\addplot [color=mycolor1,solid,line width=1.0pt,mark=square,mark options={solid},forget plot]
  table[row sep=crcr]{%
1	7.2020990892403\\
};
\addplot [color=orange!60!purple,solid,line width=1.0pt,mark=square*,mark options={solid,fill=orange!60!purple},forget plot]
  table[row sep=crcr]{%
2	8.58496250072116\\
};
\addplot [color=orange!60!purple,solid,line width=1.0pt,mark=square,mark options={solid},forget plot]
  table[row sep=crcr]{%
2	8.39746250072122\\
};
\addplot [color=mycolor2,solid,line width=1.0pt,mark=square*,mark options={solid,fill=mycolor2},forget plot]
  table[row sep=crcr]{%
3	8.92481250360578\\
};
\addplot [color=mycolor2,solid,line width=1.0pt,mark=square,mark options={solid},forget plot]
  table[row sep=crcr]{%
3	7.8554945079334\\
};
\addplot [color=mycolor2,solid,line width=1.0pt,mark=*,mark options={solid,fill=mycolor2},forget plot]
  table[row sep=crcr]{%
4	9.16992500144231\\
};
\addplot [color=mycolor2,solid,line width=1.0pt,mark=o,mark options={solid},forget plot]
  table[row sep=crcr]{%
4	8.8130693156409\\
};
\addplot [color=mycolor2,solid,line width=1.0pt,mark=diamond*,mark options={solid,fill=mycolor2},forget plot]
  table[row sep=crcr]{%
5	10.1699250014423\\
};
\addplot [color=mycolor2,solid,line width=1.0pt,mark=diamond,mark options={solid},forget plot]
  table[row sep=crcr]{%
5	8.97292313915094\\
};
\addplot [color=mycolor1,solid,line width=1.0pt,mark=*,mark options={solid,fill=mycolor1},forget plot]
  table[row sep=crcr]{%
6	11.1996723448364\\
};
\addplot [color=mycolor1,solid,line width=1.0pt,mark=o,mark options={solid},forget plot]
  table[row sep=crcr]{%
6	10.2098830694765\\
};
\addplot [color=orange!60!purple,solid,line width=1.0pt,mark=diamond*,mark options={solid,fill=orange!60!purple},forget plot]
  table[row sep=crcr]{%
7	11.4918530963297\\
};
\addplot [color=orange!60!purple,solid,line width=1.0pt,mark=diamond,mark options={solid},forget plot]
  table[row sep=crcr]{%
7	10.0404194364884\\
};
\addplot [color=orange!60!purple,solid,line width=1.0pt,mark=*,mark options={solid,fill=orange!60!purple},forget plot]
  table[row sep=crcr]{%
8	11.7548875021635\\
};
\addplot [color=orange!60!purple,solid,line width=1.0pt,mark=o,mark options={solid},forget plot]
  table[row sep=crcr]{%
8	10.7163619228065\\
};
\addplot [color=mycolor1,solid,line width=1.0pt,mark=diamond*,mark options={solid,fill=mycolor1},forget plot]
  table[row sep=crcr]{%
9	13.3987436919382\\
};
\addplot [color=mycolor1,solid,line width=1.0pt,mark=diamond,mark options={solid},forget plot]
  table[row sep=crcr]{%
9	11.6879548975824\\
};
\node[right, align=left, inner sep=0mm, font=\bfseries, text=black]
at (axis cs:9.65,14,0) {a)};
\end{axis}
\end{tikzpicture}%
		}
	\caption{Plot a) shows maximal entropy (filled markers) and the related mutual information for all device sets. Plot b) shows the $P_{total}$ (filled) and the average of power states (empty markers) from all devices.  }
	\label{fig:dataSetsHI}
\end{figure}

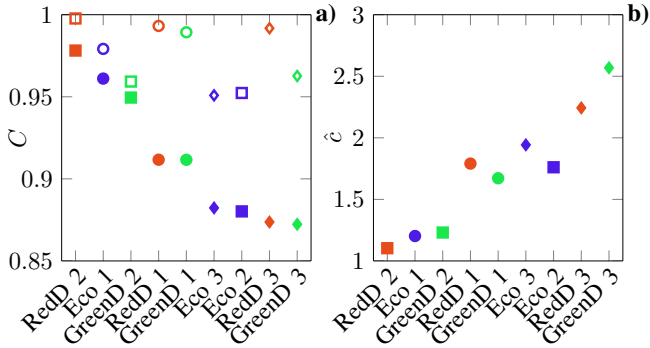
\begin{figure}[h]
	\centering
	\resizebox {\columnwidth} {!} {
%
%
\definecolor{mycolor1}{rgb}{0.30000,0.10000,0.90000}%
\definecolor{mycolor2}{rgb}{0.10000,0.90000,0.30000}%
\begin{tikzpicture}

\begin{axis}[%
width=.4\columnwidth,
height=3.5cm,
at={(.6\columnwidth,0.366667in)},
scale only axis,
clip=false,
separate axis lines,
every outer x axis line/.append style={black},
every x tick label/.append style={font=\color{black}},
xmin=0.5,
xmax=9.5,
xtick={1,2,3,4,5,6,7,8,9},
xticklabels={{RedD 2},{Eco 1},{GreenD 2},{RedD 1},{GreenD 1},{Eco 3},{Eco 2},{RedD 3},{GreenD 3}},
xticklabel style={
            inner sep=1pt,
            anchor=north east,
            rotate=45 },
every outer y axis line/.append style={black},
every y tick label/.append style={font=\color{black}},
ymin=1,
ymax=3,
ylabel={$\hat{c}$},
ylabel style={yshift=-6.5mm} 
]
\addplot [color=orange!60!purple,solid,line width=1.0pt,mark size=2pt,mark=square*,mark options={solid,fill=orange!60!purple},forget plot]
  table[row sep=crcr]{%
1	1.10344827586207\\
};
\addplot [color=mycolor1,solid,line width=1.0pt,mark size=2pt,mark=*,mark options={solid,fill=mycolor1},forget plot]
  table[row sep=crcr]{%
2	1.20250521920668\\
};
\addplot [color=mycolor2,solid,line width=1.0pt,mark size=2pt,mark=square*,mark options={solid,fill=mycolor2},forget plot]
  table[row sep=crcr]{%
3	1.23076923076923\\
};
\addplot [color=orange!60!purple,solid,line width=1.0pt,mark size=2pt,mark=*,mark options={solid,fill=orange!60!purple},forget plot]
  table[row sep=crcr]{%
4	1.79067357512953\\
};
\addplot [color=mycolor2,solid,line width=1.0pt,mark size=2pt,mark=*,mark options={solid,fill=mycolor2},forget plot]
  table[row sep=crcr]{%
5	1.67164179104478\\
};
\addplot [color=mycolor1,solid,line width=1.0pt,mark size=2pt,mark=diamond*,mark options={solid,fill=mycolor1},forget plot]
  table[row sep=crcr]{%
6	1.94266441821248\\
};
\addplot [color=mycolor1,solid,line width=1.0pt,mark size=2pt,mark=square*,mark options={solid,fill=mycolor1},forget plot]
  table[row sep=crcr]{%
7	1.76086956521739\\
};
\addplot [color=orange!60!purple,solid,line width=1.0pt,mark size=2pt,mark=diamond*,mark options={solid,fill=orange!60!purple},forget plot]
  table[row sep=crcr]{%
8	2.24299065420561\\
};
\addplot [color=mycolor2,solid,line width=1.0pt,mark size=2pt,mark=diamond*,mark options={solid,fill=mycolor2},forget plot]
  table[row sep=crcr]{%
9	2.56898192197907\\
};
\node[right, align=left, inner sep=0mm, font=\bfseries, text=black]
at (axis cs:9.65,3,0) {b)};
\end{axis}

\begin{axis}[%
width=.4\columnwidth,
height=3.5cm,
at={(.1\columnwidth,0.366667in)},
scale only axis,
clip=false,
separate axis lines,
every outer x axis line/.append style={black},
every x tick label/.append style={font=\color{black}},
xmin=0.5,
xmax=9.5,
xtick={1,2,3,4,5,6,7,8,9},
xticklabels={{RedD 2},{Eco 1},{GreenD 2},{RedD 1},{GreenD 1},{Eco 3},{Eco 2},{RedD 3},{GreenD 3}},
xticklabel style={
            inner sep=1pt,
            anchor=north east,
            rotate=45 },
every outer y axis line/.append style={black},
every y tick label/.append style={font=\color{black}},
ymin=0.85,
ymax=1,
ylabel={$C$},
ylabel style={yshift=-6mm} 
]
\addplot [color=orange!60!purple,solid,line width=1.0pt,mark size=2pt,mark=square*,mark options={solid,fill=orange!60!purple},forget plot]
  table[row sep=crcr]{%
1	0.978159485264591\\
};
\addplot [color=orange!60!purple,solid,line width=1.0pt,mark size=2pt,mark=square,mark options={solid},forget plot]
  table[row sep=crcr]{%
1	0.997703624858443\\
};
\addplot [color=mycolor1,solid,line width=1.0pt,mark size=2pt,mark=*,mark options={solid,fill=mycolor1},forget plot]
  table[row sep=crcr]{%
2	0.961084121653636\\
};
\addplot [color=mycolor1,solid,line width=1.0pt,mark size=2pt,mark=o,mark options={solid},forget plot]
  table[row sep=crcr]{%
2	0.979169058395872\\
};
\addplot [color=mycolor2,solid,line width=1.0pt,mark size=2pt,mark=square*,mark options={solid,fill=mycolor2},forget plot]
  table[row sep=crcr]{%
3	0.94952336132915\\
};
\addplot [color=mycolor2,solid,line width=1.0pt,mark size=2pt,mark=square,mark options={solid},forget plot]
  table[row sep=crcr]{%
3	0.959273232152213\\
};
\addplot [color=orange!60!purple,solid,line width=1.0pt,mark size=2pt,mark=*,mark options={solid,fill=orange!60!purple},forget plot]
  table[row sep=crcr]{%
4	0.911651593503911\\
};
\addplot [color=orange!60!purple,solid,line width=1.0pt,mark size=2pt,mark=o,mark options={solid},forget plot]
  table[row sep=crcr]{%
4	0.993158287811269\\
};
\addplot [color=mycolor2,solid,line width=1.0pt,mark size=2pt,mark=*,mark options={solid,fill=mycolor2},forget plot]
  table[row sep=crcr]{%
5	0.911623372105501\\
};
\addplot [color=mycolor2,solid,line width=1.0pt,mark size=2pt,mark=o,mark options={solid},forget plot]
  table[row sep=crcr]{%
5	0.989313166090644\\
};
\addplot [color=mycolor1,solid,line width=1.0pt,mark size=2pt,mark=diamond*,mark options={solid,fill=mycolor1},forget plot]
  table[row sep=crcr]{%
6	0.882299833860956\\
};
\addplot [color=mycolor1,solid,line width=1.0pt,mark size=2pt,mark=diamond,mark options={solid},forget plot]
  table[row sep=crcr]{%
6	0.950871418620511\\
};
\addplot [color=mycolor1,solid,line width=1.0pt,mark size=2pt,mark=square*,mark options={solid,fill=mycolor1},forget plot]
  table[row sep=crcr]{%
7	0.880185942815006\\
};
\addplot [color=mycolor1,solid,line width=1.0pt,mark size=2pt,mark=square,mark options={solid},forget plot]
  table[row sep=crcr]{%
7	0.952316009398426\\
};
\addplot [color=orange!60!purple,solid,line width=1.0pt,mark size=2pt,mark=diamond*,mark options={solid,fill=orange!60!purple},forget plot]
  table[row sep=crcr]{%
8	0.873698902372429\\
};
\addplot [color=orange!60!purple,solid,line width=1.0pt,mark size=2pt,mark=diamond,mark options={solid},forget plot]
  table[row sep=crcr]{%
8	0.991733803926311\\
};
\addplot [color=mycolor2,solid,line width=1.0pt,mark size=2pt,mark=diamond*,mark options={solid,fill=mycolor2},forget plot]
  table[row sep=crcr]{%
9	0.872317223637534\\
};
\addplot [color=mycolor2,solid,line width=1.0pt,mark size=2pt,mark=diamond,mark options={solid},forget plot]
  table[row sep=crcr]{%
9	0.962640985775383\\
};
\node[right, align=left, inner sep=0mm, font=\bfseries, text=black]
at (axis cs:9.65,1,0) {a)};
\end{axis}
\end{tikzpicture}%
		}
	\caption{The proficiency $C$ is shown in a). Filled markers show the maximal entropy case, empty ones the values for $\hat{p}=0.1$.
	Plot b) shows the average occupancy number $\hat{c}$.   
	}
	\label{fig:dataSetsC}
\end{figure}

\begin{table}[h]
	\small
	\centering
	\def\arraystretch{1.2}
	
			\begin{tabular}{cc|cccccc}
		\toprule
\multicolumn{2}{c|}{Device Set }  &	 $S$	&$M$ & $H_{max}$ & $I^P_{max}$ &$C_{max}$	& $\hat{c}$ \\ 
		\midrule
		
  \textbf{GreenD1}& {\color{grun}\CIRCLE} & 26     & 2352     &11.2& 10.21& 0.91 & 1.23\\
	\textbf{GreenD2}& {\color{grun}$\blacksquare$} &  15    &192    &7.59&  7.20& 0.95 &  1.10\\
	\textbf{GreenD3}& {\color{grun}$\blacklozenge$} &   30   & 10800   &  13.4&  11.69 &  0.87 &  1.76\\
			\midrule
	\textbf{RedD1} & {\color{rot}\CIRCLE} & 26    & 3456    &11.75& 10.72 &  0.91 &  1.18\\
	\textbf{RedD2} & {\color{rot}$\blacksquare$} & 17     &  384     &  8.59&  8.4  &  0.98  & 1.94\\
	\textbf{RedD3} & {\color{rot}$\blacklozenge$} & 24    &  2880    &11.49& 10.04 &   0.87  & 1.67\\
			\midrule
	\textbf{Eco1}& {\color{blau}\CIRCLE}  & 19    & 576    &9.17&  8.84 &   0.96 &   2.24\\
	\textbf{Eco2}& {\color{blau}$\blacksquare$}  & 17     & 486     & 8.92& 7.86 &  0.88  &  1.79\\
	\textbf{Eco3}& {\color{blau}$\blacklozenge$}  &23		& 1152   & 10.17&  8.97 &  0.88 &  2.57\\
		\bottomrule
	
		\end{tabular}	
	\caption{Several parameters for the nine sets of $N=6$ devices. Comparison is shown in the figures \ref{fig:dataSetsHI} and \ref{fig:dataSetsC}. } 
	\label{tab:dataSets}
	\end{table}
All the parameters shown in table \ref{tab:dataSets} result directly from power values of the devices of table \ref{tab:dataSetPvalues}. 
The values are presented within figure \ref{fig:dataSetsHI}, which shows the houses ranked according to their number of states, and in figure \ref{fig:dataSetsC} in which the set is sorted by descending proficiency. 
Figure \ref{fig:dataSetsHI} depicts entropy and mutual information for the maximal entropy case and characteristic power values, \ie the total set power $P_{total}$ and average device set power $P_{av}$ of the sets in kilowatt. $P_{av}$ is the expected average value when all devices are turned on. It is calculated by getting the average power for each device $ \langle P_d\rangle = 1/\hat{s}_d \sum_s{P_d^s} $ and then averaging the device set $P_{av}=1/N \sum^d{\langle P_d\rangle}$.
Statistically data sets with more states are expected to have higher total power values. 
GreenD2 and Eco3 are exceptions which leads to the conclusion that the bias of device selection can be significant. Conclusions about the number of states by the total power is inappropriate for individual device sets.
The average set power is between 50 and 75 \% of the total set power.

\begin{figure}[h]
\resizebox {\columnwidth} {!} {
%
%
\definecolor{mycolor1}{rgb}{0.30000,0.10000,0.90000}%
\definecolor{mycolor2}{rgb}{0.10000,0.90000,0.30000}%
\begin{tikzpicture}

\begin{axis}[%
width=.5\textwidth,
height=4cm,
scale only axis,
clip=false,
separate axis lines,
every outer x axis line/.append style={black},
every x tick label/.append style={font=\color{black}},
xmin=0,
xmax=1,
xlabel={$\hat{p}$},
every outer y axis line/.append style={black},
every y tick label/.append style={font=\color{black}},
ymin=0.85,
ymax=1,
ylabel={C},
ylabel style={yshift=-0.5mm} 
]
\addplot [color=orange!60!purple,solid,line width=2.0pt,mark size=1pt,mark=square*,mark options={solid,fill=orange!60!purple},forget plot]
  table[row sep=crcr]{%
0.05	0.999068265154858\\
0.1	0.997703624858443\\
0.15	0.996047611145102\\
0.2	0.994174812129019\\
0.25	0.992157516520333\\
0.3	0.990073851360003\\
0.35	0.988009021107182\\
0.4	0.986054035924027\\
0.45	0.984303024480025\\
0.5	0.982849632632326\\
0.55	0.981782953364113\\
0.6	0.981183492780869\\
0.65	0.981119692124443\\
0.7	0.981645435811068\\
0.75	0.982798857789229\\
0.8	0.984603046825344\\
0.85	0.987071640671444\\
0.9	0.990233442450737\\
0.95	0.994240140192837\\
};
\addplot [color=mycolor1,solid,line width=2.0pt,mark size=2.5pt,mark=*,mark options={solid,fill=mycolor1},forget plot]
  table[row sep=crcr]{%
0.05	0.984148703500952\\
0.1	0.979169058395872\\
0.15	0.974909132863903\\
0.2	0.971102223443928\\
0.25	0.967694817358556\\
0.3	0.964698914613791\\
0.35	0.962155512815054\\
0.4	0.960121164023414\\
0.45	0.958661539903677\\
0.5	0.957847414118544\\
0.55	0.957751359217607\\
0.6	0.958444477847561\\
0.65	0.959993049125955\\
0.7	0.962455447460695\\
0.75	0.965880208162018\\
0.8	0.970306857473805\\
0.85	0.975772846044883\\
0.9	0.982335838032093\\
0.95	0.990149069644771\\
};
\addplot [color=mycolor2,solid,line width=2.0pt,mark size=1.8pt,mark=square*,mark options={solid,fill=mycolor2},forget plot]
  table[row sep=crcr]{%
0.05	0.964484101128034\\
0.1	0.959273232152213\\
0.15	0.955821106621248\\
0.2	0.953271920140059\\
0.25	0.951339090420543\\
0.3	0.949901257108796\\
0.35	0.948905747047253\\
0.4	0.948334865212349\\
0.45	0.948191657179123\\
0.5	0.94849339684598\\
0.55	0.949268855486651\\
0.6	0.950557972268044\\
0.65	0.95241370567101\\
0.7	0.954906757510227\\
0.75	0.958135253431079\\
0.8	0.962244652077886\\
0.85	0.967472309045817\\
0.9	0.974263858990787\\
0.95	0.983671111200408\\
};
\addplot [color=orange!60!purple,solid,line width=2.0pt,mark size=2.5pt,mark=*,mark options={solid,fill=orange!60!purple},forget plot]
  table[row sep=crcr]{%
0.05	0.997106646258492\\
0.1	0.993158287811269\\
0.15	0.988372075451791\\
0.2	0.98286228559258\\
0.25	0.97674335671229\\
0.3	0.970142647720105\\
0.35	0.963204394078476\\
0.4	0.956091728123636\\
0.45	0.948988501970048\\
0.5	0.942101648218821\\
0.55	0.935664586621239\\
0.6	0.92994217769225\\
0.65	0.925237798697179\\
0.7	0.921903296484276\\
0.75	0.920353392254401\\
0.8	0.921090167634462\\
0.85	0.924759706186271\\
0.9	0.932324006132779\\
0.95	0.945723537384802\\
};
\addplot [color=mycolor2,solid,line width=2.0pt,mark size=2.5pt,mark=*,mark options={solid,fill=mycolor2},forget plot]
  table[row sep=crcr]{%
0.05	0.995245435193135\\
0.1	0.989313166090644\\
0.15	0.98257448907049\\
0.2	0.9752786276444\\
0.25	0.967650458120074\\
0.3	0.959905611000634\\
0.35	0.952256333128863\\
0.4	0.94491476998148\\
0.45	0.938095353488245\\
0.5	0.932016867241373\\
0.55	0.926904320039937\\
0.6	0.922990421144899\\
0.65	0.920516138012546\\
0.7	0.919729683354822\\
0.75	0.920883586750082\\
0.8	0.924230560920399\\
0.85	0.93002124953417\\
0.9	0.938515468634181\\
0.95	0.950079012909512\\
};
\addplot [color=mycolor1,solid,line width=2.0pt,mark size=4.3pt,mark=diamond*,mark options={solid,fill=mycolor1},forget plot]
  table[row sep=crcr]{%
0.05	0.960797824358298\\
0.1	0.950871418620511\\
0.15	0.942026941424683\\
0.2	0.93364614667529\\
0.25	0.925637132403508\\
0.3	0.918063293761235\\
0.35	0.911055659505334\\
0.4	0.90478232919707\\
0.45	0.899435369741275\\
0.5	0.895224640920675\\
0.55	0.892375243581314\\
0.6	0.891127063318858\\
0.65	0.891735373077566\\
0.7	0.894475034438991\\
0.75	0.899657558227431\\
0.8	0.907672399612511\\
0.85	0.919074397534881\\
0.9	0.934815191534047\\
0.95	0.957048012033173\\
};
\addplot [color=mycolor1,solid,line width=2.0pt,mark size=1.8pt,mark=square*,mark options={solid,fill=mycolor1},forget plot]
  table[row sep=crcr]{%
0.05	0.965768351035778\\
0.1	0.952316009398426\\
0.15	0.940056774129186\\
0.2	0.928652126398448\\
0.25	0.918120477021386\\
0.3	0.90856932517133\\
0.35	0.900136217382577\\
0.4	0.892972117652044\\
0.45	0.88723782977035\\
0.5	0.883105588017775\\
0.55	0.880763251082515\\
0.6	0.880420223208005\\
0.65	0.882314804879055\\
0.7	0.886723591885544\\
0.75	0.893976217780718\\
0.8	0.904482487740701\\
0.85	0.918784143731367\\
0.9	0.937668883280382\\
0.95	0.962559970354678\\
};
\addplot [color=orange!60!purple,solid,line width=2.0pt,mark size=4.3pt,mark=diamond*,mark options={solid,fill=orange!60!purple},forget plot]
  table[row sep=crcr]{%
0.05	0.996596938499582\\
0.1	0.991733803926311\\
0.15	0.985638285940581\\
0.2	0.978446794439909\\
0.25	0.970301368043342\\
0.3	0.961361708879177\\
0.35	0.951808697132811\\
0.4	0.941845775723572\\
0.45	0.931699706220277\\
0.5	0.92162114649486\\
0.55	0.91188513477514\\
0.6	0.902791398328901\\
0.65	0.894664479808486\\
0.7	0.887854485561617\\
0.75	0.882742326151503\\
0.8	0.879762518957253\\
0.85	0.879480232733443\\
0.9	0.882825543036571\\
0.95	0.891918303005785\\
};
\addplot [color=mycolor2,solid,line width=2.0pt,mark size=4.3pt,mark=diamond*,mark options={solid,fill=mycolor2},forget plot]
  table[row sep=crcr]{%
0.05	0.972063604523401\\
0.1	0.962640985775383\\
0.15	0.953711457945264\\
0.2	0.944880687365918\\
0.25	0.936065407243937\\
0.3	0.927282756415612\\
0.35	0.918604024848662\\
0.4	0.910141243612629\\
0.45	0.902044308451869\\
0.5	0.894503097261211\\
0.55	0.88775322784885\\
0.6	0.882085655689122\\
0.65	0.87786127264727\\
0.7	0.87553267702925\\
0.75	0.875676879814721\\
0.8	0.879047073283152\\
0.85	0.8866736635253\\
0.9	0.900148852960046\\
0.95	0.922702946031483\\
};
\end{axis}
\end{tikzpicture}%
		}
	\caption{The proficiency $C$ changes with the averaged device probability. The function is mainly related to the distribution of power values among the devices. }
	\label{fig:dataSets_Cpd}
\end{figure}
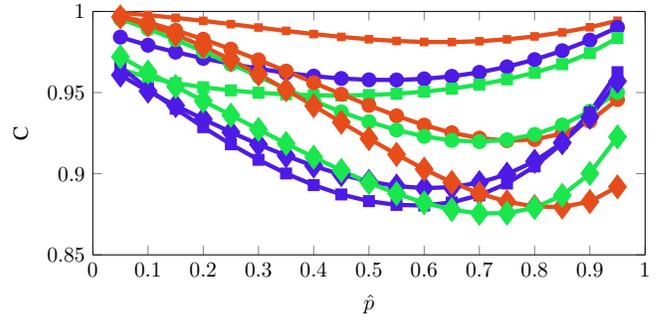

Plot a) of figure \ref{fig:dataSetsC} contains the proficiency for maximal values (filled) and the proficiency for low device probability (empty) with $\hat{p}=0.1$. The latter one is obviously closer to one and is a sample of figure \ref{fig:dataSets_Cpd}. Device usage rates have a significant influence on proficiency. 
Plot b) shows the average occupation numbers $\hat{c}$ of power states. In general it increases with decreasing proficiency. The real device sets all yield between 1 and 3, way below the values from the artificial sets A and B as listed in table \ref{tab:Bsets} and \ref{tab:AB_IandC}. 
The average occupation number measures average equality of power values. The appliance set complexity (AC) from \cite{Egarter2015} is a measure for similarity of power values (without considering their likelihood), which includes similarity.  
If the distribution of modeling- and measurement errors, which is assumed to be normal in \cite{Egarter2015} is of Delta type, the AC is expected to match the value of $\hat{c}$ for a specific device set. 
Values for AC are therefore always above $\hat{c}$.

\begin{figure}[h]
\resizebox {\columnwidth} {!} {
		\input{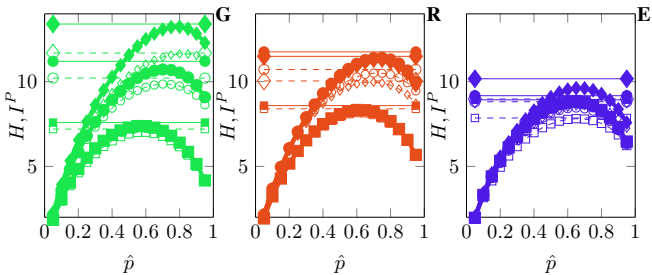}
		}
	\caption{Entropy and mutual information are shown for all the data sets, the horizontal lines mark the respective values for the maximum entropy case. 
	}
	\label{fig:dataSets_HIpd}
\end{figure}

As demonstrated in section \ref{sec:multiStates} entropy and proficiency are a function of device operation probability. Figure  \ref{fig:dataSets_HIpd} shows entropy and mutual information for each device set grouped by the three data sets GreenD, RedD and Eco. The maximal entropy values are depicted by horizontal lines.  
In figure \ref{fig:dataSets_Cpd}  the proficiency for the values presented in figure \ref{fig:dataSets_HIpd} is plotted. 
The device sets react differently to varying device probability. 
The three RedD sets get close to 1 for low $\hat{p}$ which means that the power values with view  involved devices are generally distinguishable. 
The device sets RedD3 and GreenD3 have lowest $C$-values at comparative high $\hat{p}$ which means the indistinguishable power values include many devices, making them less likely to occur. 
The GreenD2 set is special, as proficiency is barely influenced by $\hat{p}$. 
It is ranked lowest according number of states, \ie $M$ or $H_{max}$ in figure \ref{fig:dataSetsC}a, while for low $\hat{p}<0.1$ proficiency is smaller than for GreenD3, which is uppermost in figure \ref{fig:dataSetsC}a.


\section{Discussion }
\label{sec:Discussion}

Load disaggregation is the decoding process within an information communication problem. 
The code depends exclusively on device attributes and their representation in the power draw. 
Entropy, as a measure for the amount of initial states (equaling possible device configurations), has the advantage that it adds up in case of two merging device sets. 
This is generally not true for the mutual information of power values, which is an entropy type measure as well. 
The values for the maximal entropy case are a bound for more realistic cases that include the probabilities of devices to run and the power values, respectively. 
Proficiency gives the fraction of information about the device states which can be reproduced from the power values. 
It therefore might qualify as an upper bound for detection rates of NILM algorithms. 
To show to which extend this is true would require the evaluation of a NILM algorithm with a considerable set of power draws. Further it is necessary to define mutual information for continuous power values with respect to the signal to noise ratio.

A set of power draws could be used for follow up projects. 
The estimation of the single device power values by analysis of the power draws histogram would improve unsupervised NILM. 
Furthermore the assessment of device probabilities by simple measures. 
For a specific power draw the total consumed average power in relation to the total power of the device set allows to estimate the average device run times. 
The proportion of time steps without any running device indicates similar reasoning and is as easy to estimate.  
\begin{figure}[htbp]
	\centering
%
%
\begin{tikzpicture}

\begin{axis}[%
width=.18\textwidth,
height=3.5cm,
at={(.29\textwidth,0.8cm)},
scale only axis,
separate axis lines,
every outer x axis line/.append style={black},
every x tick label/.append style={font=\color{black}},
xmin=0,
xmax=1,
xlabel={$\hat{p}$},
every outer y axis line/.append style={black},
every y tick label/.append style={font=\color{black}},
ymin=0,
ymax=0.4,
ylabel={$p_0$},
ylabel style={yshift=-5mm},
legend style={legend cell align=left,align=left,draw=black}
]
\addplot [color=blue,only marks,mark=triangle,mark options={solid}]
  table[row sep=crcr]{%
0.1	0.3486784401\\
0.2	0.107374182399999\\
0.3	0.0282475248999999\\
0.4	0.00604661760000003\\
0.5	0.0009765625\\
0.6	0.00010485760000023\\
0.7	5.90490000018651e-06\\
0.8	1.02400000034919e-07\\
0.9	1.00000230318642e-10\\
};
\addlegendentry{set A};

\addplot [color=blue,only marks,mark=o,mark options={solid}]
  table[row sep=crcr]{%
0.1	0.348678440100001\\
0.2	0.1073741824\\
0.3	0.0282475249000006\\
0.4	0.00604661759999992\\
0.5	0.0009765625\\
0.6	0.000104857599999786\\
0.7	5.90490000040855e-06\\
0.8	1.02399999701852e-07\\
0.9	1.00000230318642e-10\\
};
\addlegendentry{set B};

\addplot [color=black,dotted,forget plot]
  table[row sep=crcr]{%
0.1	0.3486784401\\
0.2	0.107374182399999\\
0.3	0.0282475248999999\\
0.4	0.00604661760000003\\
0.5	0.0009765625\\
0.6	0.00010485760000023\\
0.7	5.90490000018651e-06\\
0.8	1.02400000034919e-07\\
0.9	1.00000230318642e-10\\
};
\end{axis}

\begin{axis}[%
width=.18\textwidth,
height=3.5cm,
at={(.05\textwidth,0.8cm)},
scale only axis,
separate axis lines,
every outer x axis line/.append style={black},
every x tick label/.append style={font=\color{black}},
xmin=0,
xmax=1,
xlabel={$\hat{p}$},
every outer y axis line/.append style={black},
every y tick label/.append style={font=\color{black}},
ymin=0,
ymax=1,
ylabel={$P/P_{total}$},
ylabel style={yshift=-5mm}
]
\addplot [color=blue,only marks,mark=triangle,mark options={solid},forget plot]
  table[row sep=crcr]{%
0.1	0.127272727272727\\
0.2	0.181818181818182\\
0.3	0.272727272727273\\
0.4	0.381818181818182\\
0.5	0.490909090909091\\
0.6	0.581818181818182\\
0.7	0.690909090909091\\
0.8	0.8\\
0.9	0.909090909090909\\
};
\addplot [color=blue,only marks,mark=o,mark options={solid},forget plot]
  table[row sep=crcr]{%
0.1	0.0811688311688312\\
0.2	0.149350649350649\\
0.3	0.25974025974026\\
0.4	0.399350649350649\\
0.5	0.496753246753247\\
0.6	0.600649350649351\\
0.7	0.74025974025974\\
0.8	0.857142857142857\\
0.9	0.980519480519481\\
};
\addplot [color=black,dotted,forget plot]
  table[row sep=crcr]{%
0	0\\
1	1\\
};
\end{axis}
\end{tikzpicture}%
	\label{fig:Pav&p0}
	\caption{The average device operation probability correlates with the average power of the resulting time series. The likelihood of the zero power value decreases logarithmically with increasing average device probability.}
\end{figure}
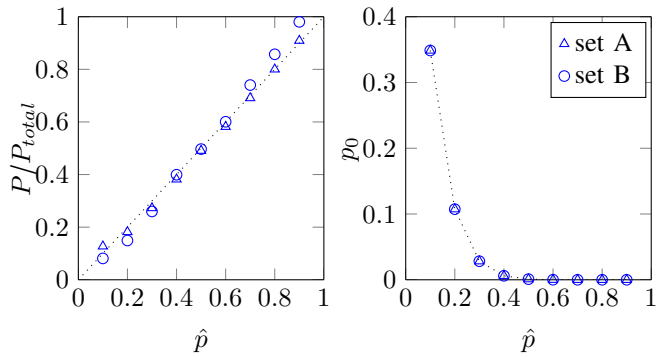

The concepts developed in this paper can be extended to any parameter space with other attributes as used in \cite{Lin2014}. Those can but do not necessarily include power values. 
The concepts help to decide if there are more promising attributes of the power draw to distinguish scenarios, or whether a single device can cause  difficulties.


\section{Summary}
\label{sec:summary}

We have modeled load disaggregation as a decoding process within an information communication problem. 
Description and improved understanding of the respective coding process helps in decoding.
If power values are used for NILM the coding scheme is likely  to be not entirely bijective as not all possible device configurations are mapped to distinguishable power values. 
We have established the calculation of entropy of initial device states, mutual information of power values and the resulting uncertainty coefficient or proficiency.  
We demonstrated that the proficiency is highly dependent on the device running probability, especially for devices with multiple values of power consumption. 
We used artificial exemplary device sets as well as real measured values of devices that were repeatedly used for other load disaggregation studies to demonstrate the meaning of these parameters. 
The insights on the coding procedure from device states to aggregated power values contributes to the improvement of existing NILM algorithms.

\bibliographystyle{IEEEtran}
\bibliography{ProficiencyofPowerValues}

\begin{thebibliography}{10}
\providecommand{\url}[1]{#1}
\csname url@samestyle\endcsname
\providecommand{\newblock}{\relax}
\providecommand{\bibinfo}[2]{#2}
\providecommand{\BIBentrySTDinterwordspacing}{\spaceskip=0pt\relax}
\providecommand{\BIBentryALTinterwordstretchfactor}{4}
\providecommand{\BIBentryALTinterwordspacing}{\spaceskip=\fontdimen2\font plus
\BIBentryALTinterwordstretchfactor\fontdimen3\font minus
  \fontdimen4\font\relax}
\providecommand{\BIBforeignlanguage}[2]{{%
\expandafter\ifx\csname l@#1\endcsname\relax
\typeout{** WARNING: IEEEtran.bst: No hyphenation pattern has been}%
\typeout{** loaded for the language `#1'. Using the pattern for}%
\typeout{** the default language instead.}%
\else
\language=\csname l@#1\endcsname
\fi
#2}}
\providecommand{\BIBdecl}{\relax}
\BIBdecl

\bibitem{Peretto2010}
L.~Peretto, ``The role of measurements in the smart grid era,''
  \emph{Instrumentation Measurement Magazine, IEEE}, vol.~13, no.~3, pp.
  22--25, June 2010.

\bibitem{Hart1992}
G.~Hart, ``{Nonintrusive appliance load monitoring},'' \emph{Proceedings of the
  IEEE}, vol.~80, no.~12, pp. 1870--1891, 1992.

\bibitem{Dong2013a}
\BIBentryALTinterwordspacing
R.~Dong, L.~Ratliff, H.~Ohlsson, and S.~S. Sastry, ``{Fundamental Limits of
  Nonintrusive Load Monitoring},'' Oct. 2013. [Online]. Available:
  \url{http://arxiv.org/abs/1310.7850}
\BIBentrySTDinterwordspacing

\bibitem{Zeifman2011}
M.~Zeifman and K.~Roth, ``Nonintrusive appliance load monitoring: Review and
  outlook,'' \emph{{IEEE} Trans. Consum. Electron.}, vol.~57, no.~1, pp. 76
  --84, February 2011.

\bibitem{Zoha2012}
A.~Zoha, A.~Gluhak, M.~A. Imran, and S.~Rajasegarar, ``Non-intrusive load
  monitoring approaches for disaggregated energy sensing: A survey,''
  \emph{Sensors}, vol.~12, no.~12, pp. 16\,838--16\,866, 2012.

\bibitem{Shaw2008}
S.~Shaw, S.~Leeb, L.~Norford, and R.~Cox, ``Nonintrusive load monitoring and
  diagnostics in power systems,'' \emph{{IEEE} Trans. Instrum. Meas.}, vol.~57,
  no.~7, pp. 1445--1454, July 2008.

\bibitem{Liang2010}
J.~Liang, S.~Ng, G.~Kendall, and J.~Cheng, ``Load signature study, part i:
  Basic concept, structure, and methodology,'' \emph{{IEEE} Trans. Power Del.},
  vol.~25, no.~2, pp. 551--560, 2010.

\bibitem{Baranski2004}
M.~Baranski and J.~Voss, ``Genetic algorithm for pattern detection in nialm
  systems,'' in \emph{Proceedings of {IEEE} International Conference on
  Systems, Man and Cybernetics}, 2004.

\bibitem{Srinivasan2006}
D.~Srinivasan, W.~S. Ng, and A.~Liew, ``Neural-network-based signature
  recognition for harmonic source identification,'' \emph{{IEEE} Trans. Power
  Del.}, vol.~21, no.~1, pp. 398--405, 2006.

\bibitem{Bier2012}
T.~Bier, D.~Abdeslam, J.~Merckle, and D.~Benyoucef, ``Smart meter systems
  detection and classification using artificial neural networks,'' in
  \emph{IECON 2012 - 38th Annual Conference on IEEE Industrial Electronics
  Society}, Oct 2012, pp. 3324--3329.

\bibitem{Xu2015}
Y.~Xu and J.~Milanovic, ``Artificial-intelligence-based methodology for load
  disaggregation at bulk supply point,'' \emph{{IEEE} Transactions on Power
  Systems}, vol.~30, no.~2, pp. 795--803, March 2015.

\bibitem{Kramer2012}
\BIBentryALTinterwordspacing
O.~Kramer, O.~Wilken, P.~Beenken, A.~Hein, A.~HÌwel, T.~Klingenberg,
  C.~Meinecke, T.~Raabe, and M.~Sonnenschein, ``\BIBforeignlanguage{English}{On
  ensemble classifiers for nonintrusive appliance load monitoring},'' in
  \emph{\BIBforeignlanguage{English}{Hybrid Artificial Intelligent Systems}},
  ser. Lecture Notes in Computer Science, E.~Corchado, V.~Snášel, A.~Abraham,
  M.~Woźniak, M.~Graña, and S.-B. Cho, Eds.\hskip 1em plus 0.5em minus
  0.4em\relax Springer Berlin Heidelberg, 2012, vol. 7208, pp. 322--331.
  [Online]. Available: \url{http://dx.doi.org/10.1007/978-3-642-28942-2\_29}
\BIBentrySTDinterwordspacing

\bibitem{Altrabalsi2014}
H.~Altrabalsi, J.~Liao, L.~Stankovic, and V.~Stankovic, ``A low-complexity
  energy disaggregation method: Performance and robustness,'' in
  \emph{Computational Intelligence Applications in Smart Grid (CIASG), 2014
  IEEE Symposium on}, Dec 2014, pp. 1--8.

\bibitem{Liao2014}
J.~Liao, G.~Elafoudi, L.~Stankovic, and V.~Stankovic, ``Non-intrusive appliance
  load monitoring using low-resolution smart meter data,'' in \emph{Proc. IEEE
  International Conference on Smart Grid Communications (SmartGridComm'14)},
  Venice, Italy, 2014.

\bibitem{gonccalves2011unsupervised}
H.~Gon{\c{c}}alves, A.~Ocneanu, M.~Berg{\'e}s, and R.~Fan, ``Unsupervised
  disaggregation of appliances using aggregated consumption data,'' \emph{The
  1st KDD Workshop on Data Mining Applications in Sustainability (SustKDD)},
  2011.

\bibitem{Zia2011}
T.~Zia, D.~Bruckner, and A.~Zaidi, ``A hidden markov model based procedure for
  identifying household electric loads,'' in \emph{Proceedings of Annual
  Conference on {IEEE} Industrial Electronics Society ({IECON})}, 2011.

\bibitem{pattem2012}
S.~Pattem, ``Unsupervised disaggregation for non-intrusive load monitoring,''
  in \emph{Machine Learning and Applications (ICMLA), 2012 11th International
  Conference on}, vol.~2, Dec 2012, pp. 515--520.

\bibitem{Egarter2014}
D.~Egarter, V.~Bhuvana, and W.~Elmenreich, ``Paldi: Online load disaggregation
  via particle filtering,'' \emph{{IEEE} Transactions on Instrumentation and
  Measurement}, vol.~64, no.~2, pp. 467--477, Feb 2015.

\bibitem{Zoha2013}
A.~Zoha, A.~Gluhak, M.~Nati, and M.~Imran, ``Low-power appliance monitoring
  using factorial hidden markov models,'' in \emph{Proceedings of {IEEE} Eighth
  International Conference on Intelligent Sensors, Sensor Networks and
  Information Processing}, 2013.

\bibitem{Zico2012}
Z.~Kolter and T.~Jaakkola, ``Approximate inference in additive factorial {HMMs}
  with application to energy disaggregation,'' in \emph{Proceedings of the
  International Conference on Artifical Intelligence and Statistics}, 2012.

\bibitem{Kim2011}
H.~Kim, M.~Marwah, M.~F. Arlitt, G.~Lyon, and J.~Han, ``{Unsupervised
  Disaggregation of Low Frequency Power Measurements},'' in \emph{Proceedings
  of the 11th {SIAM} International Conference on Data Mining}, 2011.

\bibitem{Shao2013}
H.~Shao, M.~Marwah, and N.~Ramakrishnan, ``A temporal motif mining approach to
  unsupervised energy disaggregation: Applications to residential and
  commercial buildings,'' in \emph{Proceedings of the Twenty-Seventh {AAAI}
  Conference on Artificial Intelligence, July 14-18, 2013, Bellevue,
  Washington, {USA.}}, 2013.

\bibitem{goncalves_unsupervised_2011}
H.~Goncalves, A.~Ocneanu, and M.~Berges, ``Unsupervised disaggregation of
  appliances using aggregated consumption data,'' in \emph{Proceedings of {KDD}
  Workshop on Data Mining Applications in Sustainability ({SustKDD})}, 2011.

\bibitem{Johnson2013}
M.~J. Johnson and A.~S. Willsky, ``Bayesian nonparametric hidden semi-markov
  models,'' \emph{J. Mach. Learn. Res.}, vol.~14, no.~1, pp. 673--701, Feb.
  2013.

\bibitem{Parson2014}
O.~Parson, S.~Ghosh, M.~Weal, and A.~Rogers, ``An unsupervised training method
  for non-intrusive appliance load monitoring,'' \emph{Artificial
  Intelligence}, vol. 217, no.~0, pp. 1 -- 19, 2014.

\bibitem{Cover2005}
\BIBentryALTinterwordspacing
T.~M. Cover and J.~A. Thomas, \emph{{Elements of Information Theory Wiley
  Series in Telecommunications and Signal Processing: Amazon.de: Thomas M.
  Cover, Joy A. Thomas: Fremdsprachige B\"{u}cher}}.\hskip 1em plus 0.5em minus
  0.4em\relax Wiley, 2005. [Online]. Available:
  \url{http://www.amazon.de/Elements-Information-Theory-Telecommunications-Processing/dp/0471241954}
\BIBentrySTDinterwordspacing

\bibitem{White2008}
J.~V. White, S.~Steingold, and C.~G. Fournelle, ``{Performance Metrics for
  Group-Detection Algorithms Mathematical formulation},'' in \emph{Computing
  Science and Statistics}, 2008, p.~15.

\bibitem{greend}
A.~Monacchi, D.~Egarter, W.~Elmenreich, S.~D'Alessandro, and A.~M. Tonello,
  ``{GREEND}: an energy consumption dataset of households in {I}taly and
  {A}ustria,'' in \emph{Proc. of {IEEE} International Conference on Smart Grid
  Communications (SmartGridComm)}, Venice, Italy, Nov 2014.

\bibitem{kolter2011}
J.~Z. Kolter and M.~J. Johnson, ``{REDD: A Public Data Set for Energy
  Disaggregation Research},'' in \emph{Proceeding of the SustKDD Workshop on
  Data Mining Applications in Sustainability}, 2011.

\bibitem{Beckel2014}
\BIBentryALTinterwordspacing
C.~Beckel, W.~Kleiminger, R.~Cicchetti, T.~Staake, and S.~Santini, ``The eco
  data set and the performance of non-intrusive load monitoring algorithms,''
  in \emph{Proceedings of the 1st ACM Conference on Embedded Systems for
  Energy-Efficient Buildings}, ser. BuildSys '14.\hskip 1em plus 0.5em minus
  0.4em\relax New York, NY, USA: ACM, 2014, pp. 80--89. [Online]. Available:
  \url{http://doi.acm.org/10.1145/2674061.2674064}
\BIBentrySTDinterwordspacing

\bibitem{Egarter2015}
\BIBentryALTinterwordspacing
D.~Egarter, M.~P\"{o}chacker, and W.~Elmenreich, ``{Complexity of Power Draws
  for Load Disaggregation},'' Jan. 2015. [Online]. Available:
  \url{http://arxiv.org/abs/1501.02954}
\BIBentrySTDinterwordspacing

\bibitem{Figueiredo2013}
M.~Figueiredo, B.~Ribeiro, and A.~de~Almeida, ``Electrical signal source
  separation via nonnegative tensor factorization using on site measurements in
  a smart home,'' \emph{{IEEE} Transactions on Instrumentation and
  Measurement}, vol.~PP, no.~99, pp. 1--1, 2013.

\bibitem{Lin2014}
Y.-H. Lin and M.-S. Tsai, ``Non-intrusive load monitoring by novel neuro-fuzzy
  classification considering uncertainties,'' \emph{{IEEE} Transactions on
  Smart Grid}, vol.~5, no.~5, pp. 2376--2384, Sept 2014.

\end{thebibliography}

\end{document}